\documentclass[twocolumn,superscriptaddress]{revtex4-2}
\usepackage{graphicx}
\usepackage{dcolumn}
\usepackage{bm}
\usepackage{xcolor}
\usepackage{soul}
\usepackage{amsfonts,footnote} 
\usepackage{amsmath}
\usepackage{physics,hyperref}
\usepackage{amssymb}

\begin{document}

\title{Topological photonics: fundamental concepts, recent developments, and future directions}

\author{Mahmoud Jalali Mehrabad}
\email[]{mjalalim@umd.edu}
\affiliation{Joint Quantum Institute, NIST/University of Maryland, College Park, Maryland 20742, USA}
\affiliation{Quantum Technology Center, University of Maryland, College Park, Maryland 20742, USA}

\author{Sunil Mittal}
\affiliation{Department of Electrical and Computer Engineering, Northeastern University, Boston, Massachusetts 02115, USA}

\author{Mohammad Hafezi}
\email[]{hafezi@umd.edu}
\affiliation{Joint Quantum Institute, NIST/University of Maryland, College Park, Maryland 20742, USA}
\affiliation{Quantum Technology Center, University of Maryland, College Park, Maryland 20742, USA}
\date{\today}

\begin{abstract}
Topological photonics is emerging as a new paradigm for the development of both classical and quantum photonic  architectures. What makes topological photonics remarkably intriguing is the built-in protection as well as intrinsic unidirectionality of light propagation, which originates from the robustness of global topological invariants. In this Perspective, we present an intuitive and concise pedagogical overview of fundamental concepts in topological photonics. Then, we review the recent developments of the main activity areas of this field, categorized into linear, nonlinear, and quantum regimes. For each section, we discuss both current and potential future directions, as well as remaining challenges and elusive questions regarding the implementation of topological ideas in photonics systems.
\end{abstract}

\maketitle
\tableofcontents

\section{Introduction}

\subsection{Key demonstrations in developments of topological photonics}

The entry of topology into physics started with the discovery of the quantum Hall effect in 1980 \cite{Klitzing1980}, in which Hall conductance was demonstrated to be robustly quantized in a 2D electron gas. Subsequently, it was realized that such robustness is due to the topological properties of the system energy bands \cite{Thouless1982}. The idea of band structure topology was later extended to a wider class of systems known as topological insulators \cite{Hasan2010,Qi2011}. Meanwhile, it was realized that such phenomena are not limited to  electronic systems and they can be also realized in any bosonic system. This was initially considered in the context of ultracold atoms, both in rotating Bose-Einstein condensates and optical lattices with synthetic gauge fields \cite{Cooper2019} and followed up by other bosonic systems such as photonics \cite{Ozawa2019,Price2022}, acoustics \cite{Yang2015TopologicalAcoustics}, phononics \cite{Zhang2010TopologiaclPhononics}, electronic circuits \cite{Ningyuan2015TopologicalElectronics}, and mechanics \cite{Nash2015,Huber2016TopologicalMechanics}. Specifically, in the photonic context, an analog of the quantum Hall model was proposed to realize a one-way edge state for the propagation of electromagnetic field in gyromagnetic photonic crystals \cite{Haldane2008,Raghu2008}, and subsequently demonstrated \cite{Wang2008,Wang2009}. However, to break time-reversal symmetry (TRS) this scheme relies on the presence of external magnetic fields, while the magneto-optical response of materials is weak. 

To address this issue, several theoretical proposals were put forward to synthesize magnetic fields for photons \cite{Koch2010,UmucalIlar2011,Hafezi2011,Khanikaev2012}. This was followed by two experimental demonstrations of topological edge states in optical systems without external fields \cite{Rechtsman2013,Hafezi2013}. To bring these ideas to photonic crystals,  the realization of spin \cite{Wu2015} and valley \cite{Ma2016} quantum Hall models were theoretically proposed. Subsequently, the spin Hall \cite{Barik2018} and valley Hall \cite{Shalaev2018} topological photonic crystals were experimentally demonstrated.

\begin{figure}[t!]
\includegraphics[width=0.99\columnwidth]{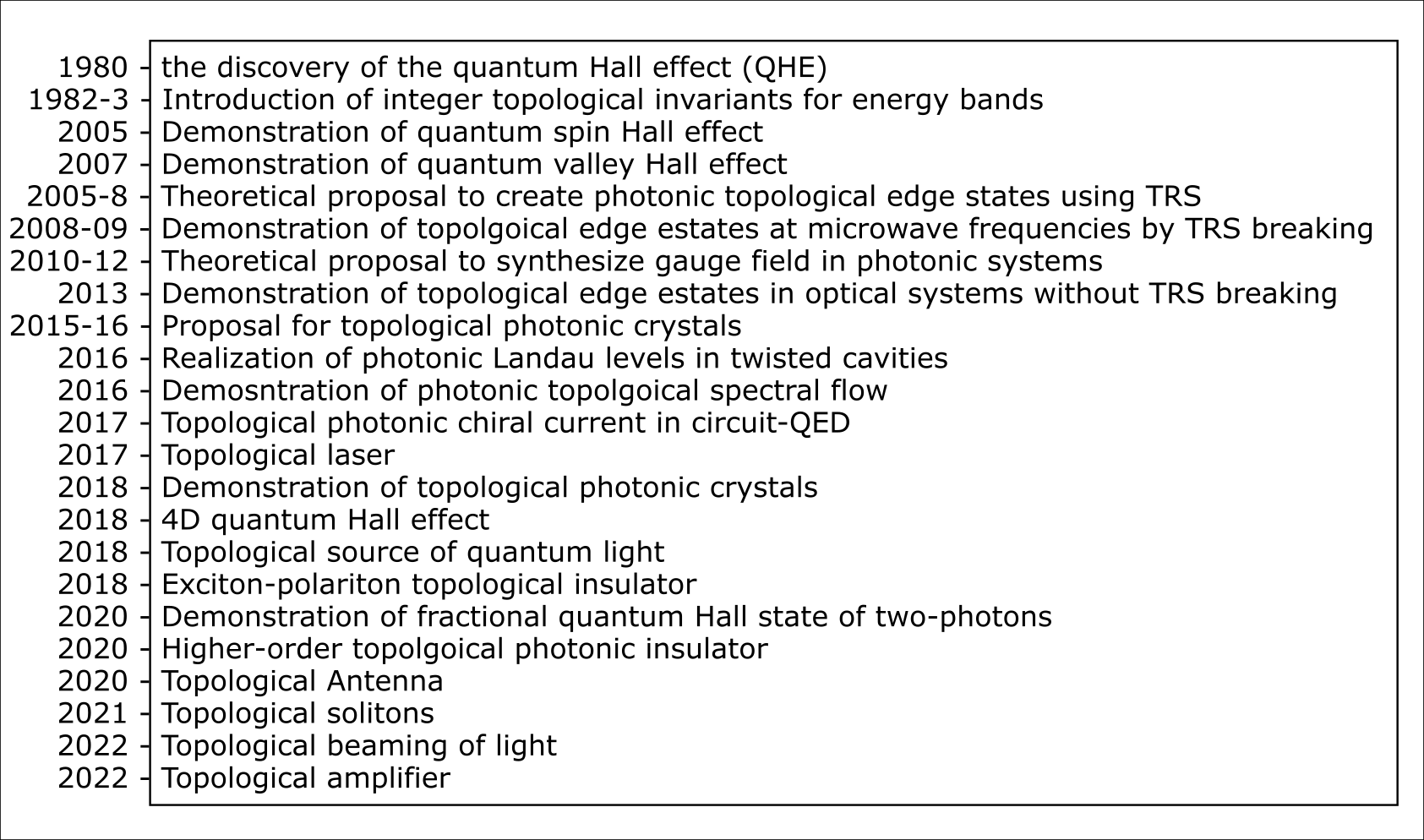}
\caption{ A selection of key developments in topological photonics (references are listed in the main text).}
\label{fig:history}
\end{figure}

However, topological invariants are not directly accessible in photonic systems. Specifically, photons are bosons, and quantization of conductance does not apply in this context. Nevertheless, quantum Hall physics can be manifested in the form of a spectral flow \cite{Laughlin1981}, which was experimentally observed in 2016 \cite{Mittal2016}. To expand the field of topological photonics into the nonlinear regime, several types of topological lasers were demonstrated \cite{Bahari2017,Bandres2018,Jean2017,Gao2020,Yang2022}. Although the nature and degree of robustness of these lasers are still subject to investigation.  Extending to the quantum regime, topological quantum sources of light were demonstrated \cite{Mittal2018,Blanco2018} around the same time. An intriguing direction is to explore strong light-matter coupling to induce strong interaction between photons. To achieve this,  integration to quantum dots \cite{Barik2018} and exciton-polariton were demonstrated in micro-cavities \cite{Klembt2018} and transition metal dichalcogenides \cite{Li2021}. Remarkably, the Laughlin state of two photons was realized \cite{Clark2020} as a major step towards few-body interacting topological systems.
Some other key developments include topological antennas \cite{Lumer2020,Gorlach2018}, 4-dimensional quantum Hall effect \cite{Zilberberg2018}, higher-order topological insulators \cite{ElHassan2019,Mittal2019,Noh2018,Chen2019,Peterson2018}, simulation of Landau levels for photons in a cavity \cite{Schine2016}, and topological solitons \cite{Mittal2021, Mukherjee2020}. Finally, three recent demonstrations showed robust topological funneling for light in a lattice geometry \cite{Weidemann2020}, photonic quantum Hall effect and generation of large orbital angular momenta \cite{Bahari2021} and topological beaming of light \cite{Lee2022}. Some of these developments have been summarized in Fig.~\ref{fig:history}.

\begin{figure}[t!]
\includegraphics[width=0.99\columnwidth]{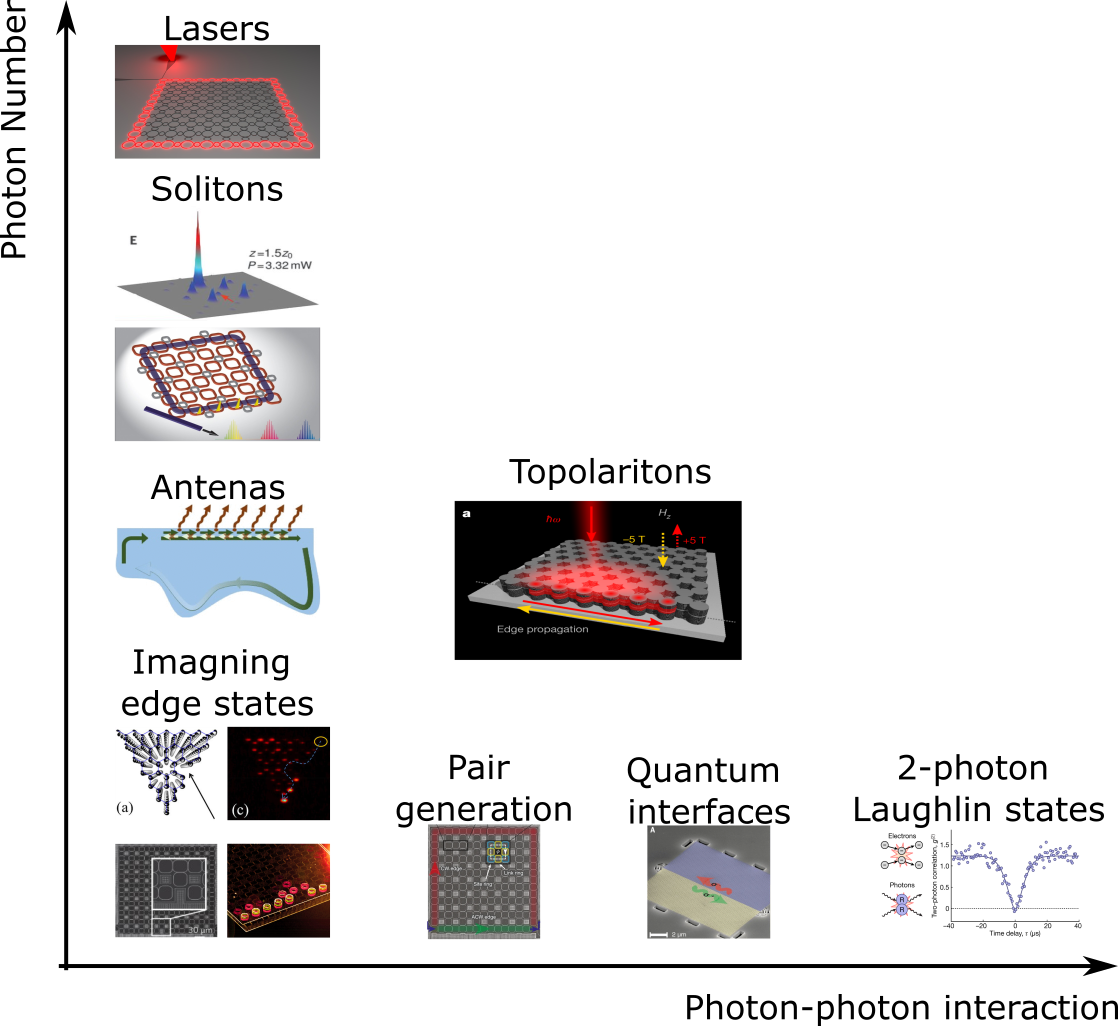}
\caption{ A selection of emerging topological photonic systems categorized based on photon number and photon-photon-interaction strength, focused on the optical and infra-red domain \cite{Bandres2018,Clark2020,Mittal2021,Klembt2018,Mukherjee2020,Barik2018,Mittal2018,Rechtsman2013,Hafezi2013,Lumer2020}.}
\label{fig:perspective}
\end{figure}

In order to have a broader perspective of the above-mentioned developments, one can classify the observed phenomena based on the involved photon number and the strength of photon-photon interaction. The classification can be seen in Fig.~\ref{fig:perspective}. In the regime of weak optical nonlinearity, classical photonic topological phenomena are shown along the vertical axis with increasing photon number, starting from low photon number cases such as silicon photonic coupled ring resonators \cite{Hafezi2013,Rechtsman2013}, to topological antennas \cite{Lumer2020,Gorlach2018}, spatial and temporal topological solitons \cite{Mittal2021,Mukherjee2020} and lasers \cite{Bahari2017,Bandres2018,Jean2017}. Moving along the horizontal axis, strong light-matter interaction enables one to induce photon-photon interaction:  Starting from the weak interaction regime (topological quantum light generation \cite{Mittal2018}, and quantum optics interface between single emitters and photonic crystals \cite{Barik2018}), to the strong interaction limit, enabling the generation of two-photon Laughlin states \cite{Clark2020}. An example of the intermediate regime of interaction and large photon number is topological polaritons in micropillar semiconductor systems \cite{Klembt2018,Amo2017}.

\subsection{Scope and aims}

The scope of this Perspective is to introduce basic concepts  and discuss recent development and potential future directions in the field of topological photonics. We hope that this Perspective is useful for researchers with no background in topological physics who are interested to explore this exciting field. 

This Perspective is structured in four sections consisting of linear, nonlinear, and quantum photonic topological systems. The linear section will include a concise pedagogical section to introduce the minimum intuitive and mathematical descriptions of the key concepts required to study topological photonics. Then, linear photonic implementations such as topological photonic crystals and passive waveguides and routers are reviewed in this section. The nonlinear section focuses on nonlinear effects in topological systems such as lasers, spatial and temporal solitons, and frequency combs. The quantum section will review the topological quantum sources of light, topological protection for the propagation of quantum states, chip-integrated quantum emitters, and systems of strongly interacting electron-photons. The last section includes detailed remarks on current challenges and more specific potential future directions as well.

It is not possible to discuss all the developments in the field of topological photonics in all platforms and frequency domains. Here, the focus of this Perspective is on the optical and infrared domain. In particular, we do not focus on the microwaves regime, for which a comprehensive recent review is available elsewhere \cite{Carusotto2020}. While we provide basic simple ideas behind linear topological photonics, we do not provide a pedagogical review of nonlinear and quantum topological photonics. We refer the reader to a comprehensive review that summarizes the development up to 2020 for nonlinear photonics \cite{Smirnova2020} and quantum \cite{Yan2021,Blanco2020}. Moreover, we refer the reader to a review on Non-Hermitian topological photonics \cite{Parto2020}, which is another emerging direction. Higher-dimension and higher-order topological photonics are reviewed in Ref.\cite{Kim2020}. A review on topological lasers can be found here \cite{Segev2020}. Synthetic dimension and other developments can be found in a recent exhaustive roadmap \cite{Price2022}.

In this Perspective, we highlight challenges for specific topological platforms, some of which are fundamental and some are more technical, as potential directions for future research. These challenges will be separately discussed in linear, nonlinear, and quantum topological photonics. 

Before proceeding, we make a clarifying remark. While the word ``topological insulator’’ has been extensively used in the literature of topological photonics, we refrain from its usage in this Perspective to avoid confusion. Strictly speaking, almost all photonic states studied so far are \textit{not} insulating states, due to their bosonic nature. In electronic systems, either because of Pauli exclusion (Fermionic nature) or interaction, such as band or Mott insulators, the system can be in an insulating state if it is probed at the corresponding Fermi levels.  In the photonic context, there is no notion of the Fermi level, since bosonic states can have unlimited occupation in the absence of interaction. Instead, one can have a \textit{photonic bandgap} and the transmission of light can be zero if photons are injected into the system within the frequency bandwidth of this bandgap.

Moreover, it is important to distinguish between general topological states and \textit{topologically-ordered states}. We use the former as a general term for any state with classical or quantum topological properties, such as vortex states, Chern band insulators. We reserve the latter term for strongly interacting systems, where the order is a consequence of interaction and entanglement, and therefore, can be defined and classified accordingly. This field is an active area of research, mainly theoretical due to the lack of clean and unambiguous experimental platforms (see \cite{Wen2017} for a recent review). With this definition, topological states encompass topologically-ordered states. In this Perspective, we mainly focus on only topological states, primarily in the single-particle/classical physics regime, and only briefly discuss topologically-ordered states.

\section{Linear topological photonics}

In the following sections, we start with a very broad introduction to the role of topology in photonic systems, and then we introduce models and relevant photonic implementation of these concepts. 

\subsection{Concept of Topological Invariants}
\begin{figure}[t!]
\includegraphics[width=0.99\linewidth]{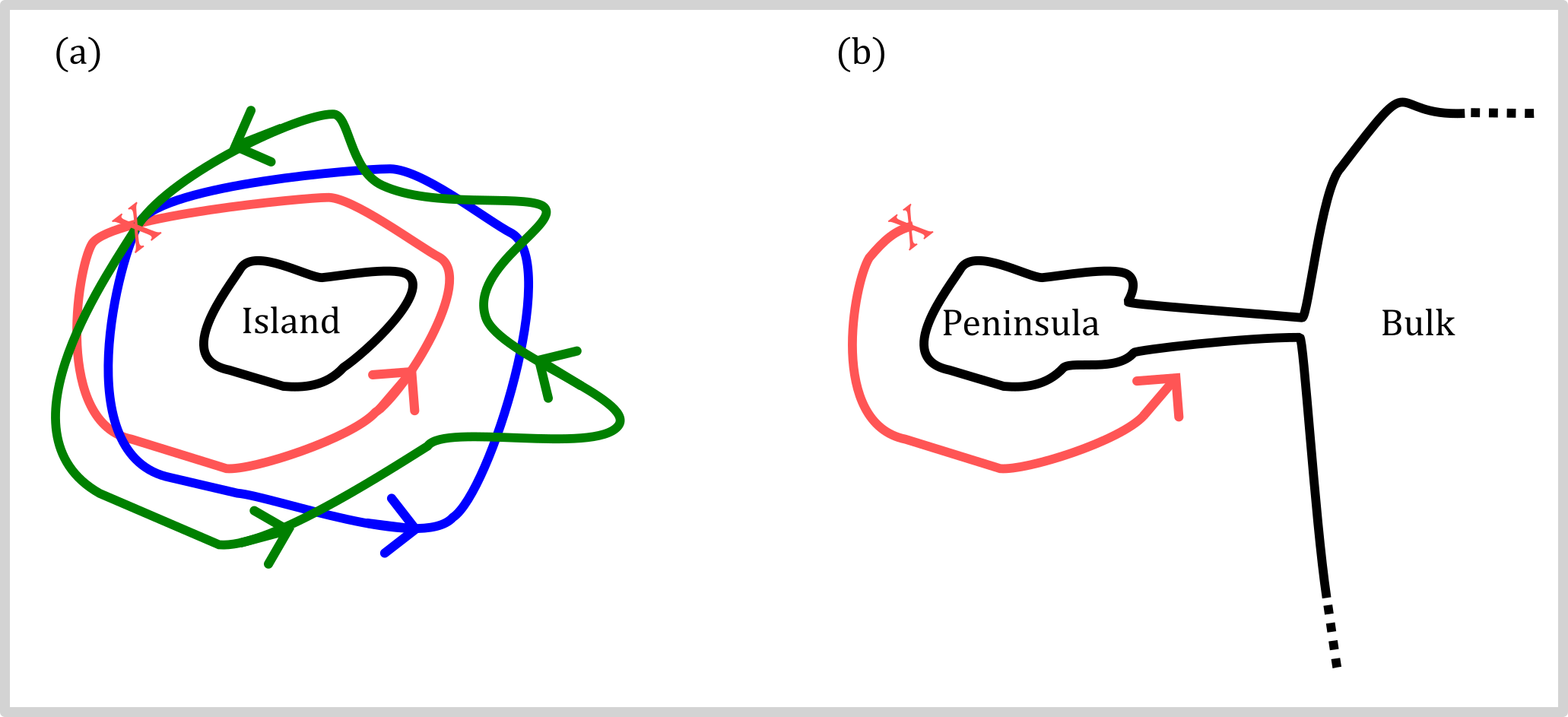}
\caption{Number of lapses when starting from a point (marked by the red cross) and going around an (a) island is an integer number, while it is zero for (b) a peninsula. Note that this integer number is invariant regarding the island's shape, swimming direction, or the taken path.}
\label{fig1}
\end{figure}

Topology is a branch of mathematics that studies the general or global characteristics of a system. For example, when studying a system of geometrical objects, instead of the specific shapes, topology primarily deals with how objects are connected. In other words, topology is concerned with the global geometrical characteristics of a system, rather than the specifics of its building blocks. As an intuitive example, one can consider the case of swimming around an \textit{island} versus a \textit{peninsula} \footnote{from Latin peninsula, from paene ‘almost’ + insula ‘island’}. Note that regardless of the shape of an island, we call it an island, but once its topology is changed, we use a different word (Figure ~\ref{fig1}).
Starting from a point and coming back to the same location, a swimmer can swim around the island, a process that does not depend on the shape of the island, or the path taken. However, the number of lapses around the island, which is an integer number (from $\mathbb{Z}$), is \emph{topologically} robust, and that number can be considered as a \textit{topological invariant}. We associate the sign of the integer to the clockwise (CW) vs the counter-clockwise (CCW) orientation of the swimmer. Note that this number is  always zero for a peninsula since a complete round trip around is not possible. The robustness here means that under small perturbations of the island's shape or the swimming path (specific ``local" features of the swimmer's path and the island's shape), the integer number (global topological property of the system) will remain invariant.
If one relates a physical observable (conduction, transmission, resistance, etc.) to such integer numbers, then that observable will be similarly topologically robust. This is one of the central motivations for the implementation of topology in physics.

The case of the island and peninsula is a classical example, with no notion of phase. A  classical wave or quantum-mechanical analog such an example can be realized by considering how electron (or photon) wavefunction winds around a certain point. An example is a vortex state in two dimensions, where the wavefunction phase can wind an integer number of times. Considering a polar coordinate $(r,\theta)$, in the presence of rotational symmetry, the wavefunction can be of the form $\psi(r,\theta)=\rho(r) e^{i m \theta}$, where the phase winds an integer number of times $m$, and the radial part of the wavefunction $\rho(r)$ has a singularity at the center. More generally, in the absence of rotational symmetry, the wavefunction can be described as $\psi(r,\theta)= \ket{\psi(r,\theta)} e^{i\phi(r,\theta)}$, where $\phi(r,\theta)$ is the local phase. Therefore, in the context of photons, these states can be thought of as a solution to the Maxwell equation, where weak spatial variation in the dielectric constant can deform the spatial form of this wavefunction but can not change the winding number $m$, i.e. $\oint {\nabla} \phi(r,\theta) \cdot  d\vec{l} =2\pi m$. This is already an example of the topological robustness of a photonic wavefunction in space.

\begin{figure}[t!]
\includegraphics[width=0.99\linewidth]{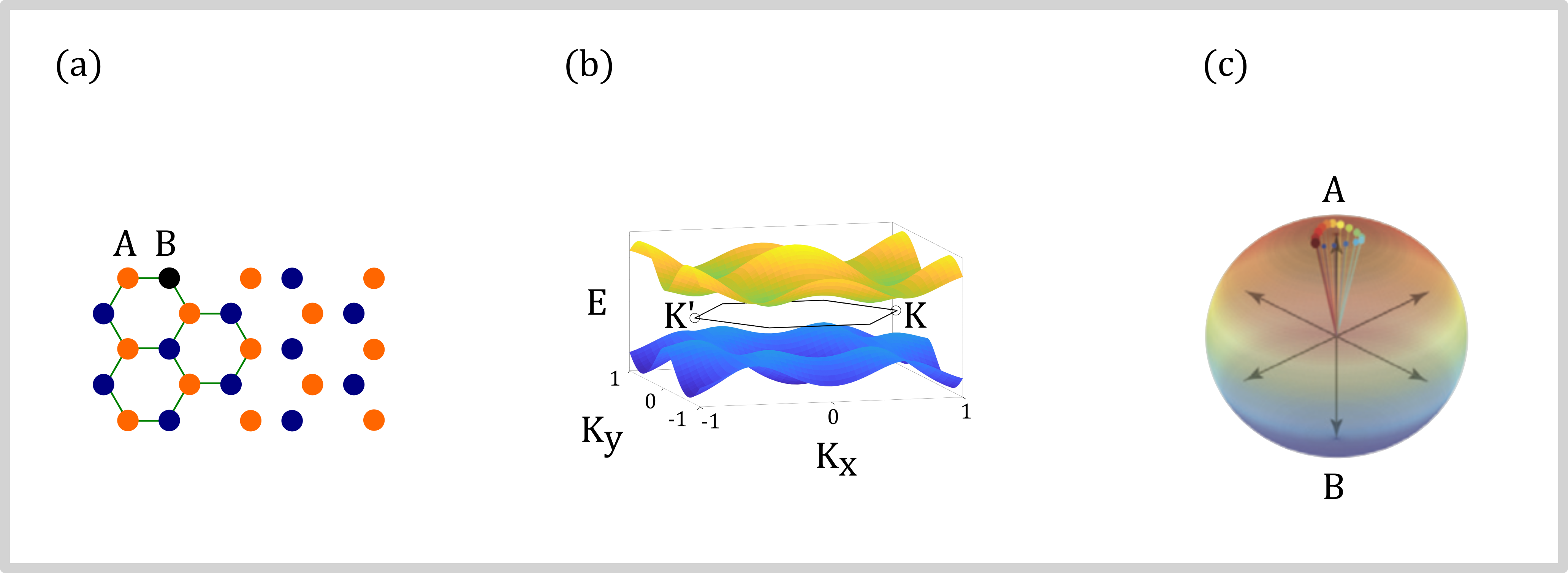}
\caption{(a) A periodic (honeycomb) photonic crystal with sub-lattices A and B. (b) Band structure of the photonic crystal, with an energy band gap separating the valence (blue) and conduction (yellow) bands. (c) Eigenvectors of the system in a unit sphere.}
\label{fig:concept_of_TPC}
\end{figure}

How can we generalize this idea? Consider a typical periodic photonic system such as a photonic crystal, e.g., a bipartite lattice as shown in Fig.~\ref{fig:concept_of_TPC}(a). Solving Maxwell's equation in such a periodic setting can provide several informative properties of the system, including the band gaps, group velocity, dispersion, etc. 

Most physical properties are indeed a \textit{local} function of the band structure $\varepsilon _\alpha(\vec{k})$, where $\varepsilon_\alpha$ is the energy and $\vec{k}$ is the wave vector in the corresponding Brillouin zone. Remarkably, there are other properties that depend on \textit{global} properties of the band structure. For example, let us take a two-band system, with wavefunctions denoted as $\psi_\uparrow(\vec{k})$ and $\psi_\downarrow(\vec{k})$, as shown in Fig.~\ref{fig:concept_of_TPC}(b). In particular, for such a two-band model, the state of the system can be represented by the unit vector on a Bloch sphere. Then, let us consider how the wavefunction varies as we move in the Brillouin zone (Fig.~\ref{fig:concept_of_TPC}(c)). If the system is topologically non-trivial, the wavefunction should accumulate a non-zero phase, when the unit vector is swept around a closed loop. Loosely speaking, there is an associated integer similar to the island and peninsula example. The latter case takes place in real space, while the former does so in momentum space.  Nevertheless, the robustness of certain global properties of the system remains warranted. More specifically, as long as the spatial variation in the susceptibility and the dielectric function is weak, this global integer remains invariant. Therefore, the associated photonic observable, such as a transmission, to this invariant remains robust to a certain disorder. We briefly clarify this connection in the photonic context in the following section.

For a survey of band topological models and a step-by-step derivation, the reader can consult \cite{Asbóth2016}. An introduction to quantum Hall physics can be found in  \cite{Girvin1999,Tong2016}.

\subsection{Toy model: charged particles in strong magnetic field}
In the following, we study a simple model of a charged particle in two dimensions, under a uniform magnetic field. While this model describes the topological physics behind the electronic quantum Hall effect, we later find it useful to synthesize similar physics in two-dimensional photonic systems. The key concepts, such as the role of gauge field, topological robustness, and topological edge state, can be understood using this simple model.

\subsubsection{Classical picture \label{sec:semiclassical}}
 
 In the classical picture, the electrons, with charge $e$ and mass $m$ undergo a cyclotron motion in the presence of an external magnetic field $\vec{B}$. Considering the Lorentz force, the dynamics of the velocity $\vec{v}$ in 2D is given by  
$m \frac{d\vec{v}}{dt}  = -e\vec{v}\times \vec{B}$. Assuming a circular motion with radius $R$, and angular velocity $\omega$, we have $m\omega^2R  = e\omega R B$. One immediately observes that the angular velocity is radius independent $\omega_c  = \frac{eB}{m}$ and $\omega_c$ is called the cyclotron frequency. Note that the center and the radius of this orbit are not constrained.

\subsubsection{Semi-classical picture}
For a semi-classical description, we can use the Bohr-Sommerfield quantization:$\oint \! p. \, dq = n\hbar$ to realize such orbits should be quantized. In fact, the integer $n$ is the phase winding number introduced earlier in our island analogy. In particular, one can set $n=1$, to find an orbit with the smallest possible radius.

\begin{figure}[t!]
\includegraphics[width=0.99\linewidth]{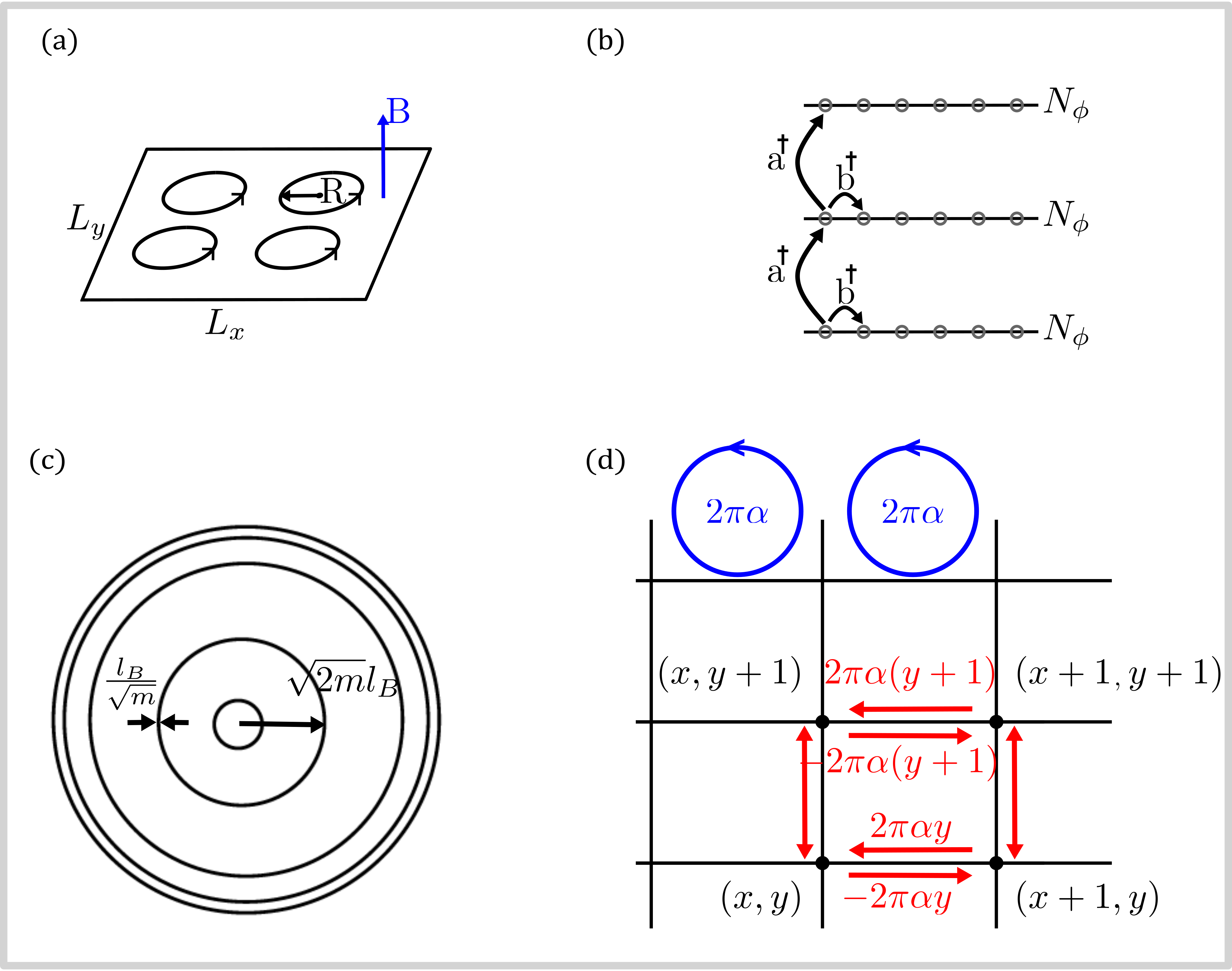}
\caption{(a) cyclotron motion of electrons with a radius of R in a 2D electron gas under a perpendicular magnetic field which generates a quantum of flux due to each electron orbit. (b) energy levels of the system, with $N_\phi$-fold degeneracy. Note that the adjacent levels are separated by $\hbar \omega_c$. (c) Wavefunction concentric orbits. (d) accumulated phase for a particle looping on the sites of a lattice.}
\label{fig:uniform_magnet}
\end{figure}

Alternatively, we can use a simple Heisenberg-limited picture, which gives us a lower limit on how small the radius of the orbits can be. Specifically, assuming $\Delta p$ and $\Delta x$ to the uncertainty in momentum and position, respectively, we have 
$\triangle x  \triangle p \simeq \hbar/2$, where for a circular motion, $\Delta x \simeq R$ and $\Delta p \simeq m\omega_c R$. Apart from a factor of two, this means the smallest orbit radius is set by

\begin{equation} \label{eq5}
l_B = \sqrt{\frac{\hbar}{m\omega_c}}  = \sqrt{\frac{\hbar}{eB}},
\end{equation}
where from now on, we call this the magnetic length. Based on the Pauli exclusion principle,  we can have only a single electron in each state. Therefore, we evaluate how many of these orbits one can fit in an area of $A = L_x L_y$, as shown in Fig.~\ref{fig:uniform_magnet}a. The total number of orbits that be fit in the system is 
 \begin{equation} \label{eq6}
\begin{split}
N_\phi  = \frac{L_x L_y}{2\pi l_B ^2}= \frac{AB}{\Phi_0}
\end{split}
\end{equation}
where $\Phi_0$ is the quantum of magnetic flux, $AB$ is the total magnetic flux, and $N_\phi$ is the total number of flux quanta. This suggests that the lowest-energy state of the system is $N_\phi$-fold degenerate.

Therefore, this suggests that as long as the number of electrons is less than $N_\phi$, they can be easily fitted into the system, if we ignore the interaction between them. But if $N_e > N_\phi$, we have to pay some energy price $\hbar \omega_c$. We can revisit this argument more precisely in the quantum picture.

Moreover, in the presence of a confining potential at the boundary of the system, the degeneracy is lifted for the state at the edge of the system. These edge states are robust against a certain amount of disorder. In other words, the states form \textit{skipping orbits} to avoid disorder, instead of reversing their directions \cite{ueta1992green}. See below for a quantum description of edge states.

\subsubsection{Quantum picture: Landau quantization}

To write the Hamiltonian of a particle with a charge ($-e$) in the presence of a classical electromagnetic field, characterized by the vector potential $\vec{A}=(A_x,A_y)$, we should simply replace all momenta by $\vec{p}\rightarrow \vec{p}+e\vec{A}$:

\begin{equation}
\begin{split}
\hat{H}  & = \frac{1}{2m} (\vec{p} + e\vec{A})^2 \\
 & = \frac{1}{2m} [(p_x + eA_x)^2 + (p_y + eA_y)^2].
\end{split}
\end{equation}
It is more common to choose the Landau gauge: $\vec{A} = -yB\vec{x}$, and obtain the $N_\phi$ degeneracy by applying the boundary conditions. Instead, we can choose the symmetric gauge:
$\vec{A}  = -\frac{yB}{2}\vec{x} + \frac{xB}{2}\vec{y}$, since the derivation is more elegant and easily generalizable. Defining mechanical momenta as 

\begin{equation} \label{eq11}
\begin{split}
\hat{\Pi}_x  & = \vec{p}_x + e\vec{A}_x, \\
\hat{\Pi}_y  & = \vec{p}_y + e\vec{A}_y,
\end{split}
\end{equation}
for any magnetic field, we have 

\begin{equation}
\begin{split}
[\hat{\Pi}_x,\hat{\Pi}_y] = -ie\hbar B.
\end{split}
\end{equation}

Therefore, if the magnetic field is uniform, by properly re-scaling these momenta operators, we can consider them as position $\hat{x}$ and momentum $\hat{p}$ operators. More interestingly, the Hamiltonian is that of a harmonic oscillator $\hat{H} = \frac{1}{2m}(\hat{\Pi}_x ^2 + \hat{\Pi}_y ^2)$. The ladder operators can be defined as:

\begin{equation}
\begin{split}
\hat{a} = \frac{1}{\sqrt{2e\hbar B}} (\hat{\Pi}_x - i\hat{\Pi}_y)
\end{split}
\end{equation}

\begin{equation}
\begin{split}
\hat{a}^\dagger = \frac{1}{\sqrt{2e\hbar B}} (\hat{\Pi}_x + i\hat{\Pi}_y).
\end{split}
\end{equation}
Consequently, $[\hat{a},\hat{a}^\dagger] = 1$ and $\hat{H}=\hbar \omega_c(\hat{a}^\dagger \hat{a}+\frac{1}{2})$, as shown in Fig.~\ref{fig:uniform_magnet}(b). This tells us that the energy levels are evenly spaced by $\hbar \omega_c$, known as Landau Levels. In order to get the Landau level degeneracy, we can similarly identify another pair of operators that commute with $\hat{H}$:

\begin{equation}
\begin{split}
\widetilde{\Pi}_x &= \vec{p}_x - e\vec{A}_x,\\
\widetilde{\Pi}_y &= \vec{p}_y - e\vec{A}_y
\end{split}
\end{equation}
and similarly, \begin{equation} \label{eq:comm_alt_momentum}
\begin{split}
[\widetilde{\Pi}_x ,\widetilde{\Pi}_y]  = ie\hbar B
\end{split}
\end{equation}
and all the other commutators are zero if we choose the symmetric gauge: $[\widetilde{\Pi}_x,\hat{\Pi}_x]   = [\widetilde{\Pi}_y ,\hat{\Pi}_y] , [\widetilde{\Pi}_x,\hat{\Pi}_y], [\widetilde{\Pi}_y ,\hat{\Pi}_x]= 0$. This gives us another harmonic oscillator, where ladder operators are $\hat{b} = \frac{1}{\sqrt{2e\hbar B}} (\hat{\Pi}_x - i\hat{\Pi}_y), \hat{b}^\dagger = \frac{1}{\sqrt{2e\hbar B}} (\hat{\Pi}_x + i\hat{\Pi}_y)$, and $[\hat{b},\hat{b}^\dagger] = 1$. As shown in Fig.\ref{fig:uniform_magnet}(b), the eigenstates are simply number states corresponding to these two harmonic oscillators: 
\begin{equation}
\begin{split}
\ket {n,m} = \frac{\hat{a}^{\dagger2} \hat{b}^{\dagger2}}{\sqrt{n!m!}} \ket{0,0}.
\end{split}
\end{equation}

The explicit form of the lowest Landau level wavefunction can be found in \cite{Girvin1999,Tong2016}. Here we make a few remarks on the properties of these wavefunctions.

\begin{itemize}
    \item {The wavefunctions are concentric orbits with average radius $r = \sqrt{2m} l_B$ and a width proportional to $\frac{l_B}{\sqrt{m}}$, as shown in Fig.~\ref{fig:uniform_magnet}(c). Therefore, as $m$ increases, the orbits become more packed, until the radius hits the system size. The maximum value of $m$ is $m_{max}  = \frac{1}{2} (\frac{d}{l_B})^2$, assuming the system to be of a disk shape with radius $d$. The expression is simply the area in units of the magnetic length, which is basically the total number of magnetic flux $N_\phi$. Therefore, we recovered the Landau level degeneracy to be $N_\phi$. More precisely, $m$ from zero to $N_\phi-1$.}

    \item This physical meaning of $\widetilde{\Pi}_x$ and $\widetilde{\Pi}_y$, in the symmetric gauge, is simply the center of orbits. Specifically $\hat{X} = -\frac{\widetilde{\Pi}_y}{eB}$, $\hat{Y} = \frac{\widetilde{\Pi}_x}{eB}$. Using Eq.~\ref{eq:comm_alt_momentum}, we find that $[\hat{X},\hat{Y}]=il_B^2$, which means we can localize the orbits both in x and y coordinates. This is essentially the same argument we had in the semi-classical picture. 

    \item Unfortunately, this model up to here can not explain the integer quantum Hall effect and one needs to add both a confining potential and disorder to the model. The introduction of the two ingredients leads to the confinement of the state in the \textit{bulk} of the system and the emergence of the \textit{edge states} at the system boundary \cite{halperin1982quantized,Hatsugai1993,Hatsugai1993_2}. In fact, this concept is more general and is known as bulk-boundary correspondence, where the bulk properties dictate the properties of the edge and vice versa.  An intuitive understanding of this concept is based on gauge invariance, either through Laughlin's argument \cite{Laughlin1981} (see \cite{Girvin1999,Tong2016} for a pedagogical presentation) or the Chern-Simons response theory \cite{wen2004quantum} (see Ref. \cite{mittal2016measurementv1} supplementary material for a derivation of this concept in photonic systems).

    \item The above orbits are also eigenstates of angular momentum: $\hat{L}_Z = i\hbar (x\partial_y - y \partial_x)$, with $\hat{L}_Z {\Psi_{LLL}}^{(m)} = m\hbar {\Psi_{LLL}}^{(m)}$. In other words, these states have a well-defined non-zero phase winding, similar to our island analogy. In the presence of weak disorder, the orbits can deform but keep their phase winding. In the strong disorder limit, the states are completely washed out.

\end{itemize}
\subsection{Hofstadter Butterfly}

So far we assumed the considered system is a 2D continuum. Let us now consider that the charged particles are confined to move on a square lattice, with lattice spacing $l_s$, in the presence of a uniform magnetic field.  We then investigate to see in what regimes these two models are equivalent.

Inspired by the Aharanov-Bohm phenomena, the essence of the magnetic field is a non-zero $2\pi\alpha$ that the particle acquires on each plaquette. Formally, one needs to modify each hopping term by the corresponding gauge field on that link. This is known as Peierls substitution $J \rightarrow J\exp\left[{\frac{ie}{\hbar} \int_{\rm link} \vec{A}.\vec{dr}}\right]$, where $J$ is the hopping rate. So naturally, different gauge conventions correspond to spreading the total phase $2\pi\alpha$ along each link of a plaquette (Fig. \ref{fig:uniform_magnet}(d)). Using the Landau gauge: $(A_x,A_y) = (-B y l_s,0)$ where $(x,y)\in \mathbb{Z}$, and are simply locations on a $(N_x,Ny)$ square lattice. The Hamiltonian describing the dynamics is

\begin{equation} \label{eq:hofstadter}
\begin{split}
\hat{H} & = -J\sum \hat{a}_{x+1,y}^\dagger {\hat{a}_{x,y}}^{-2i\pi \alpha y}  
+ \hat{a}_{x,y}^\dagger {\hat{a}_{x+1,y}}^{2i\pi \alpha y} \\
& + \hat{a}_{x,y+1}^\dagger \hat{a}_{x,y} + \hat{a}_{x,y}^\dagger \hat{a}_{x,y+1},
\end{split}
\end{equation}
where $\hat{a}_{x,y}$ is the annihilation operator of particle at site $(x,y)$. So far, we are considering a single particle so the statistics of particles are not important. But in the following sections, we assume the operators obey the bosonic commutation relations.

One can verify that the total accumulated phase for a counter-clockwise propagation on a single plaquette is: $-2\pi\alpha y + 0 + 2\pi\alpha (y+1) + 0  = 2\pi\alpha$. The number of magnetic flux in each plaquette is: \begin{equation}
    \frac{\Phi}{\Phi_0}=\frac{eB l_s^2}{h} = \alpha.
\end{equation}

In other words, $\alpha$ is the fraction of a flux in a plaquette. Note that the essence is the presence of this phase, and the charge $e$, $\hbar$, etc. drop out. Therefore, one can generalize this model to neutral particles (atoms/photons) by "synthesizing" the phase. This is the key insight in Ref.~\cite{Hafezi2011}. 

\begin{figure}[t!]
\includegraphics[width=0.99\linewidth]{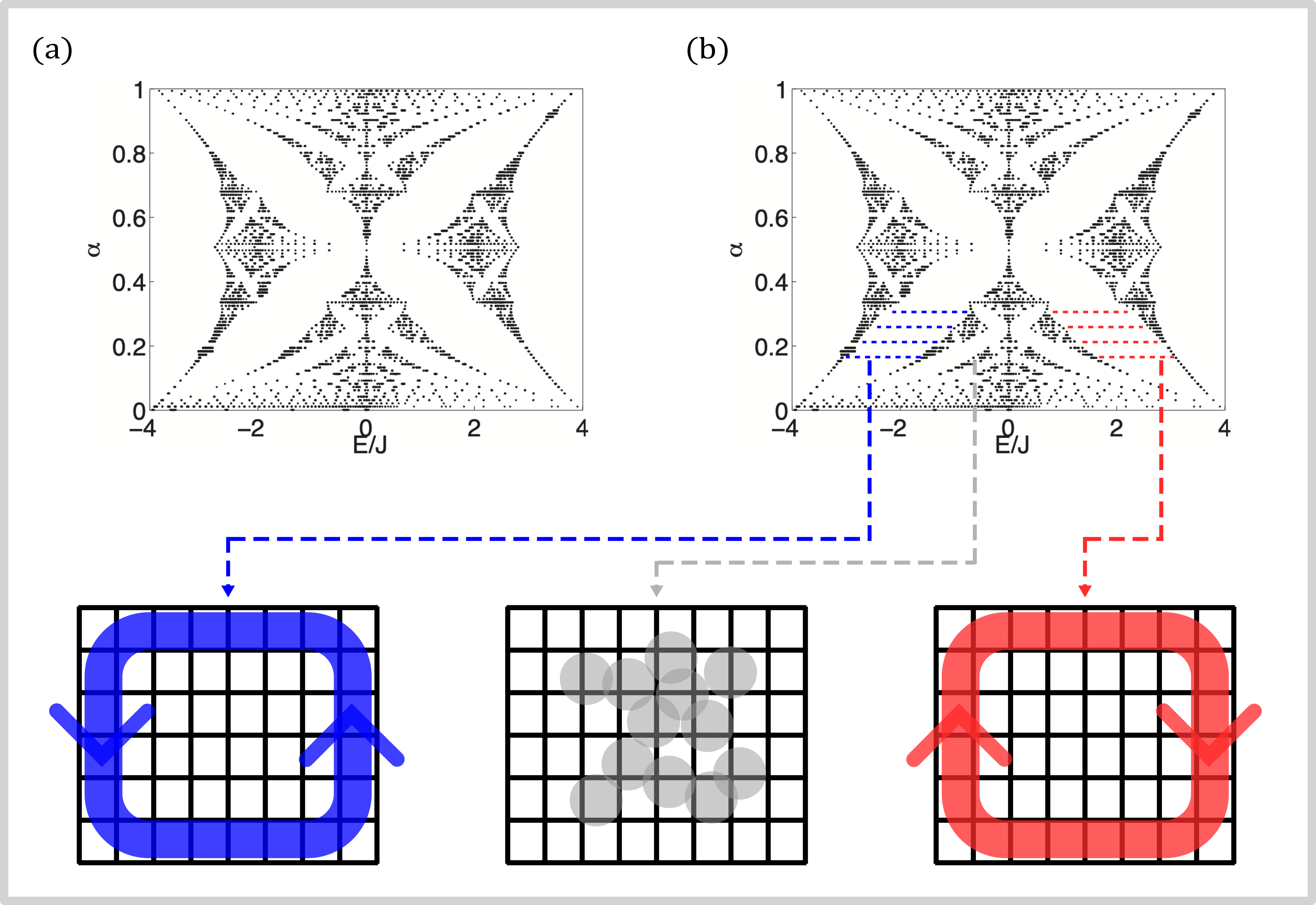}
\caption{(a,b) The spectrum of Eq.(\ref{eq:hofstadter}) on a $10\times10$ lattice, for a closed and open boundary condition, respectively. The lower panel illustrates light intensity on the lattice corresponding to three typical states. Apart from the localized bulk states in the middle, in the open boundary case, edge states can form, while  propagating in a clockwise (red) and counter-clockwise (blue) fashion.}
\label{fig:hoftstadter}
\end{figure}

The spectrum of this Hamiltonian is periodic when $\alpha \rightarrow \alpha +1$, and is known as Hofstadter Butterfly \cite{Hofstadter1976}, with many interesting fractal properties, as shown in Fig.~\ref{fig:hoftstadter}(a). One key point is the presence of band gaps when the system is considered on periodic boundary conditions, i.e., torus. Let us recall a tight-binding 2D square lattice ($\alpha = 0$ in Eq.\ref{eq:hofstadter}) of the size $(N_x,N_y)$ has a single band with $N_xN_y$ states with energies:

\begin{equation}
E(n,m)  = -2J[\cos(k_x a) + \cos(k_y a)],\end{equation} 
where $k_x l_s N_x  = 2\pi n$,$ k_y l_s N_y  = 2\pi m$. In this model, the translational symmetry is clearly broken, but if $\alpha = \frac{p}{q}$, and we go around $q$ plaquette, we get $2\pi p$ phase, which is like having no phase and a zero $\alpha$ 2D tight-binding model. This suggests that at $\frac{p}{q}$ we have $q$ bands, each containing $\frac{N_x N_y}{q}$ states, as shown in Fig.~\ref{fig:hoftstadter}(a). 

When an open boundary condition is considered, in-gap states appear, as shown in Fig.~\ref{fig:hoftstadter}(b). Such states are localized at the boundary and propagate in a chiral fashion.

Now, we can investigate the effect of the disorder. Imagine a disorder in the form of an onsite potential $\hat{a}_{x,y}^\dagger \hat{a}_{x,y}$. When the chiral edge state encounters such an obstacle, it is energetically preferred to go around the disorder site, instead of reversing the propagation path. Recall that CW and CCW edge states have different energies. Loosely speaking, this is similar to a ``quantum swimmer'', where, in the presence of an obstacle, the path is modified to make sure the wavefunction remains single-valued, instead of reversing the path.

\subsection{Photonic lattice}

\begin{figure}[t!]
\includegraphics[width=0.99\linewidth]{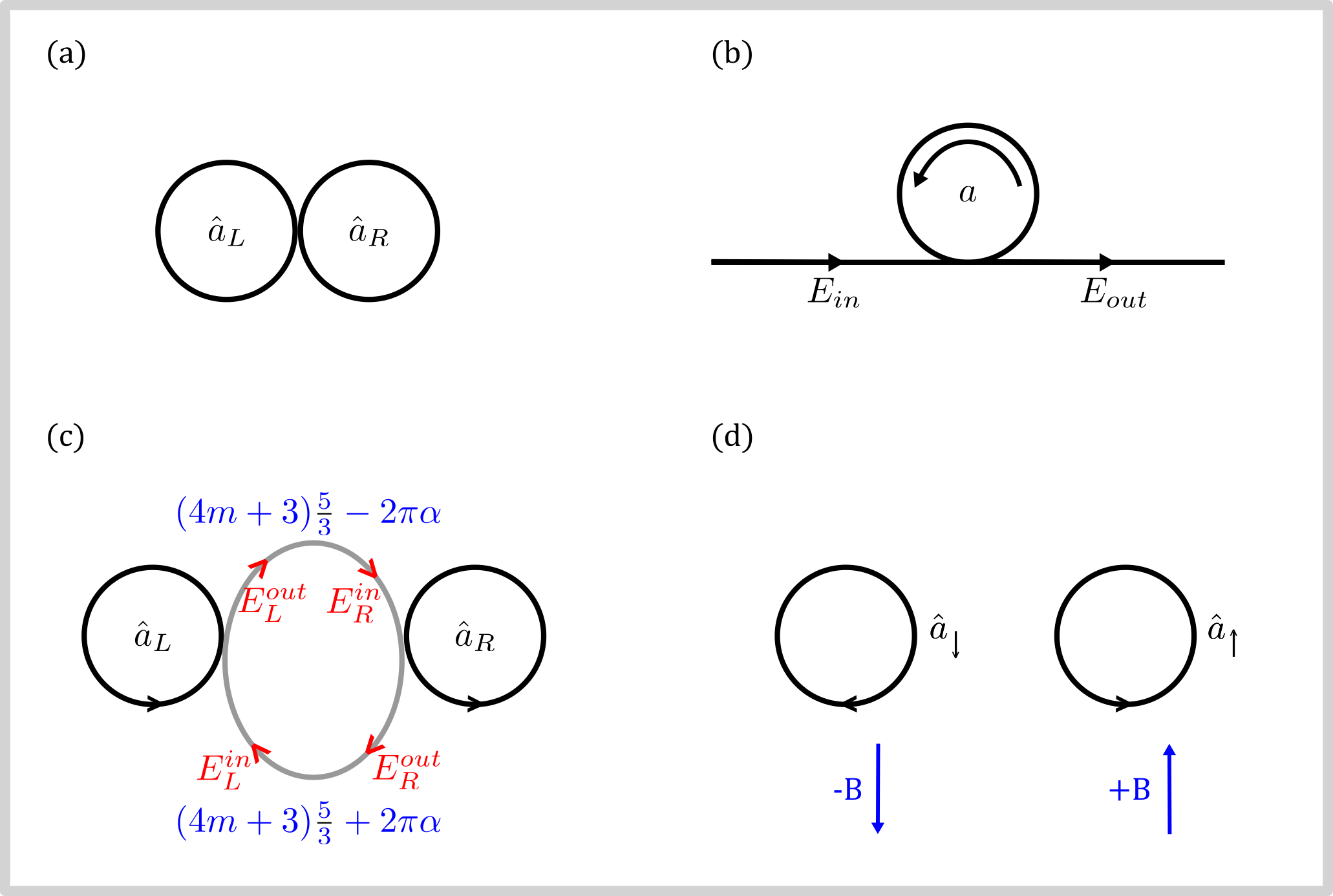}
\caption{(a) A pair of coupled ring resonators described by Eq. 14 (b) a ring resonator coupled to a waveguide that can be described by the input-out formalism described in the text. (c) two resonators coupled to each other using another resonator that is anti-resonator with the side resonators. By positioning the middle resonator and creating a differential optical path one gets Eq. 17. (d) Clockwise (left) and counter-clockwise (right) propagation under opposite magnetic fields, as illustrated by the blue arrows.}
\label{fig:coupled_rings}
\end{figure}

Can we engineer a 2D array of optical resonators to simulate the previous Hamiltonian? As we observed above, the essence is the extra phase in hopping. Let's start with two coupled resonators:

\begin{equation}
\begin{split}
\hat{H} = -J{\hat{a}_L}^\dagger \hat{a}_R -J{\hat{a}_R}^\dagger {\hat{a}_L}
\end{split}
\end{equation}
where $\hat{a}_L,\hat{a}_R$ are the annihilation operators of a photon in left and right resonators, respectively, as shown in Fig.~\ref{fig:coupled_rings}(a).  $J$ is the coupling strength and depends on the overlap of electromagnetic modes in the left and right resonators. The sign of $J$ here depends on the definition of ${\hat{a}_{L,R}}$ modes. Recall that phases usually do not have a meaning until they are evaluated for a closed loop. More importantly, two coupled resonators can not have a complex hopping phase, and our goal is to engineer one (see below Eq.\ref{eq10}).

The form is simply that of the coupled mode theory. In fact, in the absence of nonlinearity, single photon and coherent state dynamics are the same $(\hat{a} \rightarrow \langle a\rangle$) and we remove the hats going forward.

For example, the Heisenberg picture dynamics is equivalent to two coupled mode equations of motions: $\dot{a} _L= i[{H},{a} _L]= iJ a_R$, and $ \dot{a}_R  = i J a_L$.  Now, we want to consider two resonators coupled with a waveguide in between. One can use the transfer matrix formalism (see supplementary of \cite{Hafezi2013}). Here we use the quantum input-output formalism \cite{Gardiner1985} that is shorter and provides insight, following Ref.~\cite{Hafezi2013}. For a  resonator mode ($a$) coupled to a waveguide, as shown in Fig.~\ref{fig:coupled_rings}(b), decay/input can be described by: $ \dot{a}  = -\kappa a - \sqrt{2k}E^{in} $ where $E^{in},E^{out}$ is the input/output fields, respectively. The boundary condition is written as: $ E^{out}  = E^{in} + \sqrt{2\kappa} a$. We want to engineer a situation where:

\begin{equation} \label{eq10}
\begin{split}
H & = -Je^{i\phi}{a^\dagger_L} a_R  -Je^{-i\phi}{a^\dagger_R} a_L.
\end{split}
\end{equation}

Note that we can not put arbitrary coefficients in front of the two terms, the Hamiltonian should be Hermitian. Consider the scheme in Fig.~\ref{fig:coupled_rings})c, where two resonators are coupled through  an ``anti-resonant" resonator in an asymmetric way:

The total optical length of the middle ring is chosen such that photons resonant with left/right resonators do not interfere constructively in the middle, and therefore only circulate one on the above/below arm. The total optical path is $(4m+3)\pi$, where $m$ is a positive integer. Under this condition, photons spend most of their time on the left or right resonator, and we can find an effective Hamiltonian with $\hat{a}_L$ and $\hat{a}_R$, without the middle ring. The equations of motion take the following form:

\begin{equation}
\begin{split}
E^{\rm in}_R & = E^{\rm out}_L e^{2i\pi m +i\frac{3\pi}{2} - 2\pi i\alpha} \\
& = -iE^{\rm out}_L e^{- 2\pi i\alpha}, \\
E^{\rm in}_L & = -iE^{\rm out}_R e^{+2\pi i\alpha}, \\
E^{\rm out}_{R,L} & = E^{\rm {in}}_{R,L} + \sqrt{2\kappa} a_{R,L},\\
\dot{a}_{R,L} & =  -\kappa a_{R,L} - \sqrt{2\kappa }a_{R,L} E^{\rm in}_{R,L}.
\end{split}
\end{equation}

By eliminating $E_{R,L}^{in,out}$, we find the effective Hamiltonian:

\begin{equation}
\begin{split}
H & =  -\kappa {a_{R}}^\dagger a_L e^{-2\pi i\alpha} -k{a_{L}}^\dagger a_R e^{2\pi i\alpha} 
\end{split}
\end{equation}

By choosing the length of the resonators accordingly, e.g. by increasing the $\alpha$ linearly in row number $y$, we can implement the Hofstader Hamiltonian.

If the system is driven with photons corresponding to the frequencies of the edge band (see Fig.~\ref{fig:hoftstadter}), they circulate around the system either in CW or CCW way. In other words, photons experience an effective magnetic field $B$, and orbiting around the system in CW/CCW fashion leads to opposite energies, in direct analogy to $-\vec{L} \times \vec{B}$, where $\vec{L}$ is the angular momentum. 

We have not applied any external magnetic field that breaks the time-reversal symmetry (for example this is needed in an optical isolator). However, we have a magnetic field-like Hamiltonian for a passive system. How is that possible? In fact, we have two pseudo-spins $\frac{1}{2}$ corresponding to CW/CCW circulating photons, \textit{inside each resonator}, each experiencing an opposite magnetic field. Therefore, a more accurate analogy is spin-orbit interaction, where each spin orientation experiences an opposite magnetic field $\vec{S} \times \vec{L}$ where $\vec{S}$ is that photon pseudo-spin of photons. In other words, the TRS is preserved for the entire system, but we can selectively drive the system in  a ``spin-polarized" way, e.g. pumping the CW mode of the resonator.  As long as photons do not get scattered from CW to CCW mode, each experience an opposite magnetic field.  

Since the word \textit{chiral} is preserved for edge state with broken time-reversal symmetry, here we use \textit{helical} edge states.

In any physical realization, such scattering is present, however, if the rate of such backscattering processes is slower than the hopping rate, then we can ignore such processes.   In the optics language, we need to operate in an unresolved mode coupling regime.

We treated the left/right rings as single-mode resonators, while the middle ring was treated as a waveguide. Is that correct? Yes, but this is only valid for photons close to the resonance of left/right rings. For a rigorous derivation, one needs to use the transfer matrix theory. 

We again emphasize that the Hamiltonian and the second quantization formalism are not necessary for this part. However, this formalism allows one to understand its physics without getting lost in the details of transfer matrix theory.

\subsection{Various topological photonic models and their implementations}
The above model was implemented in arrays of coupled ring resonators fabricated on $SiO_2$ operating at 1550nm wavelength \cite{Hafezi2013}. Generally, it is crucial to study how the topological invariants are manifested in such systems and what are the physical observables compared to electronic systems. For electrons, the system is filled up to the Fermi level, and then the electrical conductance is measured as the main physical observable. If the system has a non-zero integer topological invariant, the conduction is quantized, genetically with the same integer.  Filling up the Fermi sea is simply a consequence of the Pauli exclusion principle that is absent for photons. In these photonic systems, however, one can probe the system with an incoming laser field with a given frequency. If the field is resonant with any of the system's modes, the photons enter (couple into) the lattice. Otherwise, the light is completely reflected \cite{Hafezi2013}. This state spectroscopy can be used to measure topological invariants as a spectral flow when the system is subject to an extra magnetic flux \cite{hafezi2014measuring,Mittal2016}, in an analogy to Laughlin flux insertion argument \cite{Laughlin1981}.

From a more general point of view, during the development of topological photonics, several models have been developed and subject to intense research, starting from integer quantum Hall, followed by anomalous quantum Hall (also known as the Haldane model), and subsequently, spin and valley-Hall effects.
Other topological models have been also implemented in rings, for example, Su–Schrieffer–Heeger (SSH) model \cite{Smirnova2020}, and topological laser arrays \cite{Harari2018}. Other models have been also implemented in helical Floquet waveguides \cite{Rechtsman2013}. A broad overview of these models, their characteristics, and implementations  can be found here \cite{Smirnova2020}.

It is also important to note highlight here that, different from its electronic counterpart, topological photonics can offer an opportunity to harness several unique degrees of freedom which are either in part or completely unavailable in electronic systems. For example, various polarization \cite{Barik2018} and orbital angular momentum \cite{Bahari2021} degrees of freedom of light can offer powerful design flexibility and novel functionalities. For example, synthetic modal dimensions have been recently implemented to synthetic hybrid spatial-modal lattice configurations beyond conventional lattice geometries \cite{lustig2019photonic}. A review of the relevant concepts and recent implementations can be found here \cite{lustig2021topological}.
%
\subsection{Topological photonic crystals}

\begin{figure}[t!]
\includegraphics[width=0.99\columnwidth]{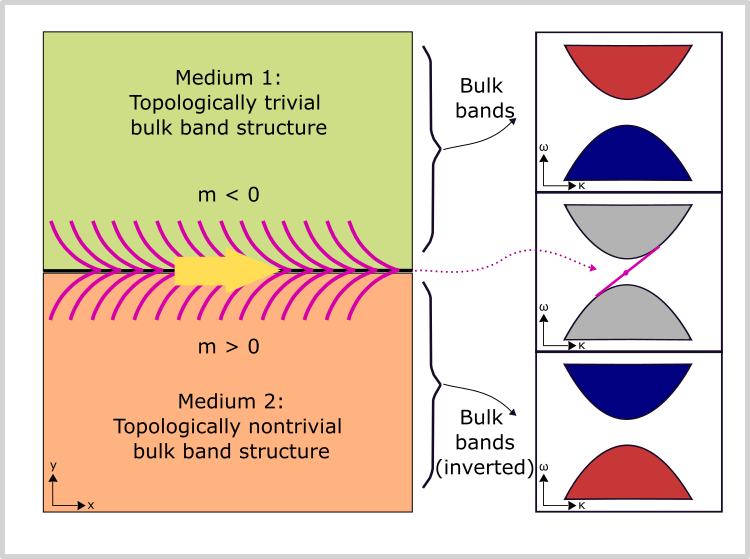}
\caption{ Illustration of formation of in-gap edge states at the interface between two topologically distinct mediums with inverted band structure. Adapted from \cite{MJmehrabad2021,Chen2018}.}
\label{fig:formation}
\end{figure}

A simple and useful topological model can be formalized for photonic crystals based on band inversion and the formation of bound states. A useful way to think about this is to use continuum models, in particular the ones that led to the emergence of the topological insulators. The essence of these models is  captured in the Jackiw-Rebbi (JR) model and the concept of band inversion. Considering a 2D system with the following dispersion:

\begin{equation*}
    \Bigl[-i\hbar v (-\sigma_x\partial_x + \sigma_y\partial_y)+m\sigma_z \Bigr] \Psi = E\Psi,
\end{equation*}
where $\Psi (x,y)$ is a spinor. $v$ and $m$ are the velocities and the effective mass, respectively. In contrast to the  electron's spin states, here the spinor represents a pseudo-spin, e.g., two modes of the electric field. We assume the mass changes sign at the crossing point $y=0$, specifically, $m(x,y)=m(y)$, and $m(0)=0$ and $\frac{dm}{dy} < 0$. There is a bound state solution at $y=0$ that propagates along the $x$-axis, described by  \begin{equation*}
    \Psi(x,y)=\frac{1}{\sqrt{2}}\begin{pmatrix} 1\\ 1 \end{pmatrix}e^{\frac{1}{\hbar v}\int_{0}^{y} m(y^\prime) \,dy^\prime}e^{ik_x x},
\end{equation*}
which is schematically shown  in Fig. ~\ref{fig:formation}. Note that if TRS is not broken, e.g., in the valley and spin-Hall effects, we get two copies of the above Hamiltonian that are connected by TR. Therefore, we have helical states (instead of chiral) that propagate in opposite directions with opposite spin (polarization for photons) \cite{Kim2020}. Here, for one of the polarization states, $m(y)$ goes from a negative to a positive sign, while it changes sign in an opposite manner for the other polarization. A more detailed investigation of these concepts can be found here \cite{Košata2021}.

An exciting implementation of the JR model is engineering topologically distinct photonic crystals (TPCs). Specifically, by changing the photonic structure, one can engineer a band inversion between two topologically distinct photonic crystals to form propagating states at the interface, as first proposed in Ref.~\cite{Wu2015} and demonstrated in Ref.\cite{Barik2018}. These states have three main characteristics: they are unidirectional (photons with opposite polarization travel in opposite directions), spatially confined (in $y$ for an interface along the x-axis)) at the boundary, and robust against certain disorders. In the spin-Hall TPCs, as described in Ref. \cite{Wu2015,Barik2016}, the opposite circular polarisation of the two edge modes can be described by the in-plane electric field profiles of the TPC's hexagonal unit cell. The in-plane electric field is highly circularly polarized, with opposite handedness for these two Jackiw-Rebbi solutions. The in-plane electric field circular polarization of ${\sigma }_{\pm }$ can be considered as pseudo-spins for this topological photonic crystal.

Similarly, one can exploit the valley degree of freedom, and engineer a band inversion. This leads to valley-Hall TPCs as first proposed in \cite{Ma2016} and demonstrated in \cite{Shalaev2018}. Similar to spin-Hall TPCs, the in-plane electric field of the TPC's unit cell has two circular polarization that propagates in opposite directions, and forms two helical topological edge states.

Topological edge states in TPCs were imaged directly in a lattice of silicon Mie resonators \cite{Peng2019}, where the opening of photonic gaps around double degenerate Dirac cone as well as the formation of topological edge states was demonstrated using high-resolution optical microscopy.  Another development was recently reported in which valley and spin degrees of freedom were shown to be presented simultaneously in a topological crystal \cite{Košata2021}.

Next, we will discuss some of the recent implementations of these types of TPCs in a variety of passive linear systems.

\begin{figure}[t!]
\includegraphics[width=0.99\columnwidth]{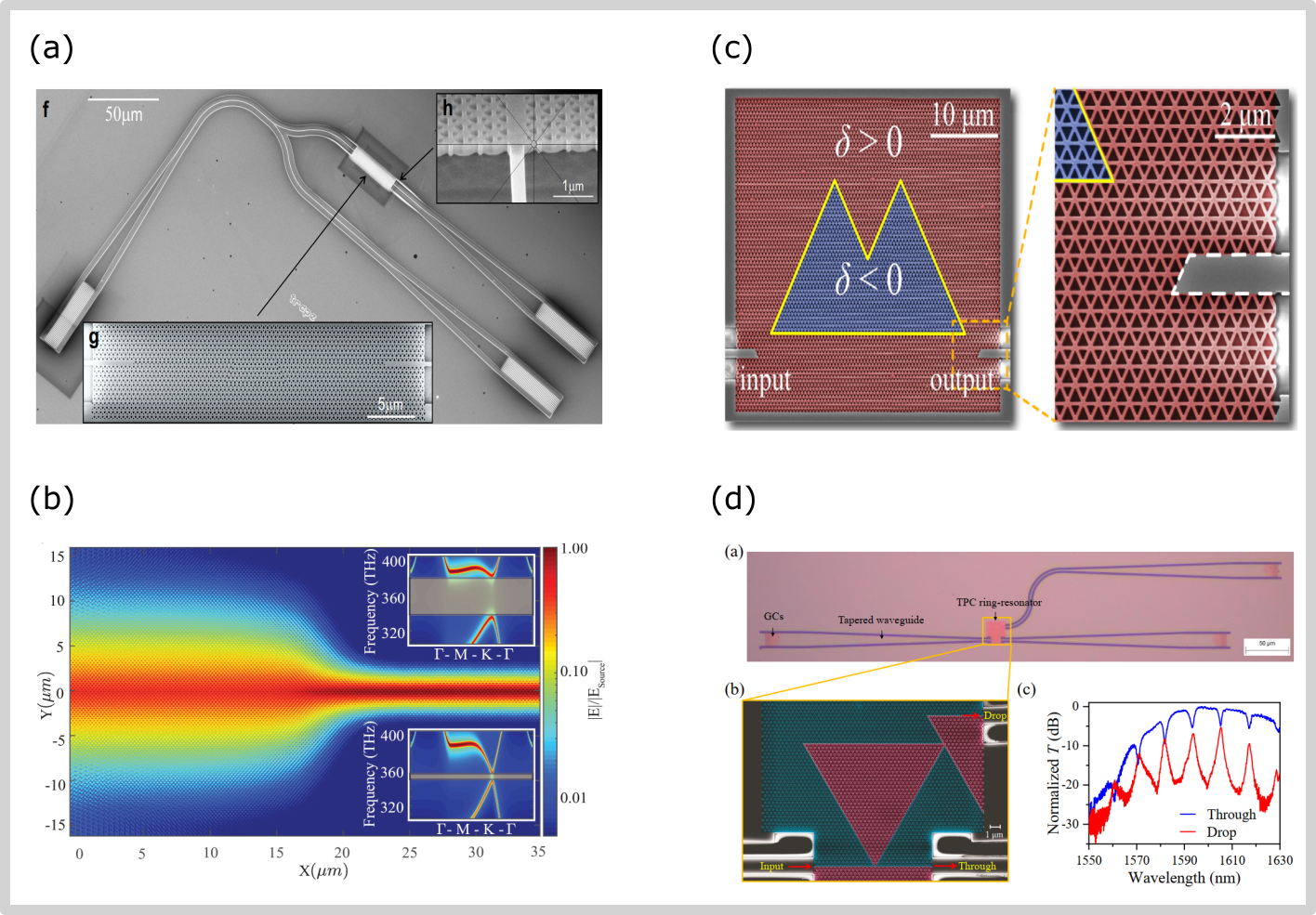}
\caption{ Implementation of topological photonic edge states in the linear regime. (a) a passive suspended valley-Hall photonic crystal waveguide \cite{Shalaev2018}. (b) a topological photonic mode taper \cite{Flower2022}. (c) a valley-Hall topological ring resonator with access bus waveguide, forming an all-pass filter \cite{Ma2019}. (d) A topological add-drop filter, comprising a valley-Hall resonator and access waveguides \cite{Gu2021}.}
\label{fig:linear}
\end{figure}

Examples of experimental demonstrations of linear topological photonic platforms are shown in ~\ref{fig:linear}. These systems include robust photonic waveguides and ring resonators in both all-pass and add-drop filter configurations. These devices use valley-Hall edge states. The quantum photonic section will cover the first demonstration of spin-Hall type photonic crystal waveguide \cite{Barik2018}.
Recently, it was proposed that adiabatic tuning of the topological bandgap in valley-Hall type photonic crystal can be utilized to form a topological mode-taper \cite{Flower2022}. Moreover, a similar approach has been implemented to realize topological rainbow trapping in photonic crystals \cite{Lu2021,Lu2022}.

\section{Nonlinear Topological Photonics}

Until very recently photonic systems (as well as other systems like acoustics, electrical circuits, etc) have been largely used to emulate single-particle electronic topological Hamiltonians, that is, systems where interactions between particles are negligible. This includes topological Hamiltonians such as the SSH model, the integer and anomalous quantum Hall effect, the spin and valley-Hall physics, higher-order topological insulators, Floquet topological insulators, and so on. Nevertheless, electronic topological systems also include effects such as the fractional quantum Hall effect where interactions between particles lead to a very rich physics. It is, therefore, natural to ask if one can use photonic topological systems to emulate "interacting" topological systems. Though single-photon interactions are very weak, we can still achieve "mean-field" nonlinear interactions between photons, at high enough photon flux, by using a nonlinear medium (whose polarization is a nonlinear function of the applied electric field). Examples of such nonlinear interactions include self-phase modulation, cross-phase modulation, sum, difference and harmonic-frequency generation, optical parametric oscillation, lasing, etc. Along these lines, one of the research directions explores if such nonlinear interactions affect the topology of the system: can they induce topological phase transitions, or are the topological edge states stable in the presence of such nonlinear interactions? On a more fundamental level, such nonlinear topological photonic systems have no counterparts in fermionic systems and can lead to the emergence of topological models that are unique to photons. In parallel, another research direction explores the applications of topological phenomena, like edge states, to engineer nonlinear processes in a medium, for example, for efficient and robust lasers, generation of quantum states of light, optical frequency conversion, etc. Even more so, one can also achieve true single-photon-level nonlinearities mediated by atoms or artificial atoms like quantum dots or superconducting qubits, or excitons in semiconductors. Such systems can then realize a photonic analog of interacting topological systems such as the fractional quantum Hall effect. In the following, we will review advances in these sub-fields of nonlinear topological photonics. We will limit our discussion to parametric nonlinearities like the Kerr effect.

\subsection{Nonlinearity Induced Topological Phase Transitions}

Optical nonlinearities, like the Kerr effect, change the refractive index of a medium as a function of the optical intensity \cite{Boyd2003}. This refractive index change can lead to a change in the on-site potential or the coupling strength between waveguides or resonators and subsequently be used to induce topological phase transitions. One of the first demonstrations of such a topological phase transition was carried out in topo-electric circuits that realized the 1D SSH model \cite{Hadad2018}. Here the nonlinearity modified the couplings between alternate lattice sites and a topologically trivial phase at low intensities transitioned to a topological phase at high enough intensities.

At optical frequencies, a nonlinearity-induced topological phase transition was proposed by Leykam et al. in a 2D coupled ring resonator system that implemented the Haldane-like anomalous quantum Hall model in a bipartite lattice \cite{Leykam2018}. In this model, a topological phase transition can be introduced by adding unequal on-site potentials $M$ (mass terms) to the two sets of lattice sites such that $M > 2J$. Leykam et al. considered a system with built-in (during fabrication) on-site potentials just below the transition threshold. A broadband high-intensity pump pulse was then injected into the link rings of the lattice such that their resonance frequency would red shift compared to the site rings. This relative frequency shift would reduce the effective coupling between the site rings and thereby, induce the topological phase transition.

An experimental realization of the nonlinearity-induced topological phase transition was demonstrated recently by Maczewsky et al. in a 2D coupled waveguide array implementing anomalous Floquet topological model \cite{Maczewsky2020}. The array is fabricated such that alternating waveguides have a non-zero on-site potential (introduced by alternating the waveguide width), and the system is topological only at each coupling region between the waveguides, the power transfer $t > 50\%$. In the linear regime, the presence of on-site potential ensures that this power transfer ratio is less than $50\%$ and the system is topologically trivial. Nevertheless, on injecting high enough power into the waveguide with a thinner core (lower effective refractive index), the Kerr nonlinearity reduces the on-site potential difference between the waveguides and increases the coupling ratio such that the system transitions to a topological phase. Note that this pump is injected only into a single waveguide and this topological phase transition is local, meaning injecting a weaker beam elsewhere in the lattice would still experience a topologically trivial phase.

\subsection{Spatial Solitons}

An optical beam propagating through a medium with optical nonlinearities, like the Kerr effect, can experience self-focusing wherein the central high-intensity region of the beam sees a higher refractive index compared to its low-intensity tails. At specific beam intensity, the self-focusing effect can exactly balance the diffraction of light, and lead to the formation of spatial solitons \cite{Chiao1964, Segev1992, Segev1998, Stegeman1999}. Optical spatial solitons have been observed in many platforms, including coupled waveguide arrays - very similar to those used for the realization of photonic topological insulators \cite{Eisenberg2002}. This immediately leads to the question: can such photonic topological systems host spatial solitons? Will these solitons live on the edge of the bulk of the system? Are these solitons robust against disorders?

\begin{figure*}
 \centering
 \includegraphics[width=0.7\textwidth]{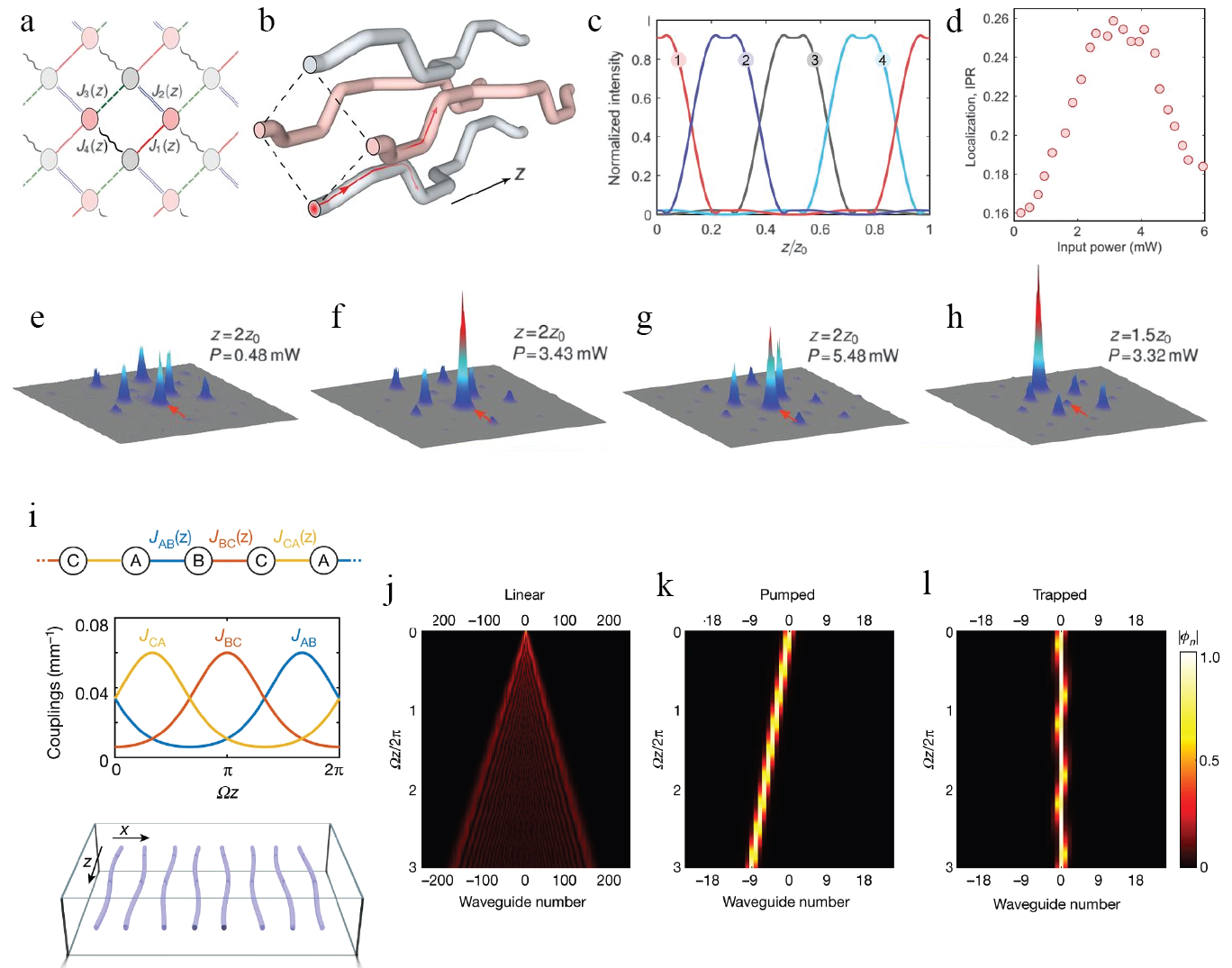}
 \caption{\textbf(a-c) Schematic of the 2D array of coupled waveguides and variation of their coupling strengths, used to observe spatial topological bulk solitons. (e-h) Intensity profile in the array showing soliton behavior. (i) Schematic of the 1D array of coupled waveguides used for pumping of topological solitons and the variation of the coupling strength between the waveguides. (j) Topological pumping in the linear regime. (k) Topological pumping of solitons at higher pump powers, and (l) trapping of solitons at very high pump powers. }
 \label{fig:spatial}
\end{figure*}

Very recently, spatial solitons were observed in topological waveguide arrays, in both bulk and edge states. Specifically, Mukherjee et al. observed bulk solitons in a 2D anomalous Floquet topological insulator \cite{Mukherjee2020}, as shown in Fig. \ref{fig:spatial}. The system consisted of a 2D array of waveguides with periodic/cyclic couplings to their nearest neighbors. On exiting the bulk waveguides at high enough input optical power, Mukherjee et al. observed solitons that undergo cyclotron motion while hopping neighboring waveguides. Because of this cyclotron motion, the intensity distribution of the soliton would repeat only after propagating through a complete period (along the waveguide) of the lattice. Nevertheless, as expected, the soliton would not diffract into the bulk of the lattice. Furthermore, the quasi-energies of the solitons were observed to be in the bandgap and the extent of localization of the solitons was observed to increase (decrease) with the increasing (decreasing) separation between the quasi-energies of the soliton and the linear band.

In another related article, Mukherjee et al. also observed a soliton-like solution on the edge states of the anomalous Floquet topological waveguide array \cite{Mukherjee2020}. As before, the input light is coupled to a single waveguide, but now on the edge of the array. Because the input is confined to a single waveguide, it can excite all the edge modes with different quasi-energies (in this case quasi-energy is propagation constant/momentum along the waveguides). The finite curvature of the edge band dispersion can then lead to the broadening of the edge excitation. Note that, even in linear topological systems, the excitations on the edge states stay localized to edge states. As such, the broadening here refers to the increase in the number of waveguides on the edge that are occupied by the beam as it propagates along the edge of the array. In the presence of nonlinearity, Mukherjee et al. observed minimal broadening - an indication of balancing the broadening of the beam against nonlinearity-induced self-focusing. Nevertheless, these soliton-like features on the edge were observed to scatter some power into the bulk of the lattice.

Following these observations of bulk and edge spatial solitons, another fascinating observation by $\text{J}\ddot{\text{u}}\text{rgensen}$ et al. has been the demonstration of Thouless pumping of solitons \cite{Jurgensen2021, Jurgensen2022, Mostaan2022}. For this experiment, they used a 1D array of coupled waveguides that simulates the off-diagonal Aubry-Andre-Harper (AAH) model for photons. In this array, the coupling strength (off-diagonal elements of the Hamiltonian matrix) between the waveguides varies periodically as a function of position along the waveguide length. This 1D model is related to the 2D Chern insulator model and exhibits an identical band structure. At the input of this array, a soliton, which is an eigenstate of the nonlinear Hamiltonian, was injected. Then, the Thouless pumping manifested as a quantized displacement of the soliton to neighboring waveguides by one unit cell after propagating one period of coupling-strength modulation. Even more, by choosing a soliton solution that bifurcated from a different band (with Chern number +2), the authors also observed quantized displacement by two unit cells in one period of propagation length. Evidently, the displacement of the soliton corresponded to the Chern number of the band from which the soliton bifurcated. Using this same platform the authors have also recently demonstrated nonlinear pumping of solitons by fractional integers that are quantized \cite{Jurgensen2023}.

In another very different platform, that of cavity polaritons, Pernet et al. also observed spatial solitons \cite{Pernet2022}. Their system consisted of a 1D array of micropillars, each of which hosts a cavity polariton. The coupling between the neighboring micropillars was staggered to realize the 1D SSH model. The nonlinearity in this system originates from the coulomb repulsion between the excitonic part of the cavity polaritons. When the topological edge state at the interface between topologically trivial and no-trivial regions was strongly pumped, a topological gap soliton was observed to be localized at the same interface state. Evidently, this topological soliton bifurcates from the mid-gap topological edge state and exhibits a spatial intensity distribution (localized on a single sublattice) that is similar to the linear case. More interestingly, when a dimer in the bulk of the array was pumped, with pump frequency in the topological bandgap, the authors observed the formation of topological bulk solitons. Furthermore, the pump power threshold for the formation of topological bulk solitons was found to be robust against defects (introduced by another laser) only in one sublattice and not in the other sublattice. Going further, the authors demonstrated that by controlling the phase of the pump excitation over two micropillars of the dimer, they could achieve sub-lattice-polarized topological solitons such that the soliton wavefunction was predominantly localized to only the sublattices, and this polarization could be controlled by controlling the relative phase of the two pump beams.

\subsection{Dissipative Kerr Temporal Solitons and Frequency Combs}

The presence of optical Kerr nonlinearity in optical resonators with multiple free spectral ranges (FSRs) can lead to the fascinating physics of temporal dissipative Kerr solitons and optical frequency combs \cite{Udem2002, Cundiff2003, Kippenberg2011, Kippenberg2018, Pasquazi2018, Gaeta2019, Diddams2020, Suh2016}. Because of the spontaneous four-wave mixing process mediated by the Kerr nonlinearity, a continuous-wave pump beam with a frequency near one of the resonances leads to the generation of new frequencies in the resonator. The energy and momentum conservation dictates that the newly generated frequencies are also close to the resonator frequencies at other FSRs. In the limit of a weak pump, this process is spontaneous and as we discussed earlier, is used to generate energy-time entangled photon pairs. With increasing pump power, the newly generated frequencies beat with the pump and also with other frequencies, ultimately leading to a stimulated four-wave mixing process. Because the generated frequencies are almost aligned with the ring resonances, they constitute an optical frequency comb with frequency separation almost equal to the FSR of the resonator. However, the comb lines are, in general, not phase-locked and the frequency comb is chaotic. In this regime, the field profile inside the resonator is also chaotic

By appropriately designing the dispersion of the resonator, mostly to be in the anomalous regime, and by tuning the pump frequency and power, it is indeed possible to get to a regime where the linear dispersion of the ring is exactly canceled by the dispersion introduced by the Kerr nonlinearity and the resonator loss is balanced by the four-wave mixing (FWM) gain. This dual balance leads to the generation of a coherent optical frequency comb where all the comb lines are phase-locked and precisely equally spaced by the FSR of the resonator. The corresponding field profile inside the resonator corresponds to one or more soliton pulses in time, called Dissipative Kerr Solitons, that propagate without dispersion. Such DKSs have been observed in a variety of resonator geometries, including whispering gallery mode, bottle, integrated ring, and Fabry-Perot resonators, and also in a number of material platforms, including silica, silicon, silicon-nitride, aluminum-nitride, silicon-carbide, and so on. From an application perspective, coherent optical frequency combs find a number of applications, for example, in precision time-keeping, spectroscopy, WDM transceivers, LiDARs, etc.

Recently, there has been growing interest in using coupled-resonator systems to engineer novel DKS solutions and comb spectra that are not accessible using single resonator geometries \cite{Tikan2021, Komagata2021, Tikan2022}. On a more fundamental level, these systems also explore the self-synchronization of coupled resonators. Some of the early demonstrations in this regard used resonators made of fiber loops or a fiber-loop coupled to an integrated ring resonator \cite{Bao2019,Xue2019}.  More recently, the field of frequency combs has seen an influx of ideas from the field of topological photonics.

\begin{figure*}
 \centering
 \includegraphics[width=0.7\textwidth]{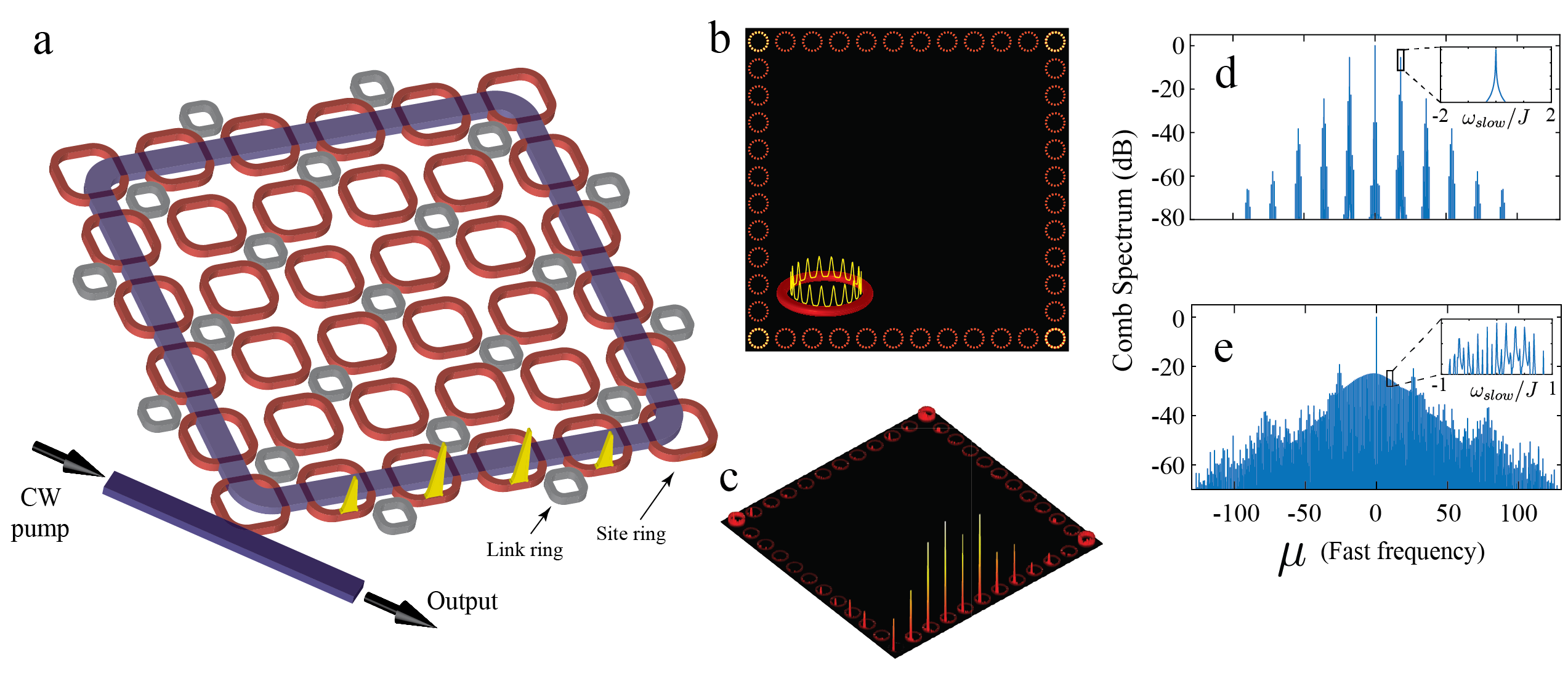}
 \caption{\textbf(a) Schematic of the 2D array of ring resonators used to generate temporal nested solitons. The resonator array simulates the anomalous quantum Hall model for photons. The pump laser is coupled to the array using the input-output waveguide. The nested comb output is collected using the same waveguide. (b) Phase-locked Turing rolls along the edge of the 2D array. (c) Phase-locked nested solitons propagating along the edge of the array. Note that there is a single soliton in each ring resonator that is a part of the nested soliton. (d) Comb spectrum in the regime of phase-locked Turing rolls showing oscillation of a single edge mode in each FSR. (e) Comb spectrum in the regime of nested solitons showing the oscillation of multiple edge modes in each FSR.}
 \label{fig:temporal}
\end{figure*}

Specifically, Mittal et al. theoretically studied the generation of DKSs and optical frequency combs in two-dimensional ring resonator arrays that, as we discussed earlier, create a synthetic magnetic field for photons, and thereby, stimulate the integer or the anomalous quantum Hall physics for photons \cite{Mittal2021b}. Given that this system realizes one copy of the anomalous Hall model near each of the single-ring resonance frequencies, it is effectively a three-dimensional system with two real and one synthetic dimension in frequency. For a linear system, the different copies at different ring resonance frequencies are uncoupled. However, the introduction of a four-wave mixing process (Kerr nonlinearity) couples these copies by mediating the hopping of photons between them. This demonstration is summarised in Fig. \ref{fig:temporal}.

As we discussed earlier, the linear dispersion and the spatial confinement of topological edge states lead to efficient phase-matching of the spontaneous four-wave mixing process for the generation of entangled photon pairs. A similar phenomenon was observed for the generation of optical frequency combs in topological ring resonator arrays. The comb generation was efficient only when the pump beam was close to one of the edge mode resonances. This can be easily understood considering that the topological edge states circulate around the complete periphery of the lattice, and hence, realize a super-ring resonator composed of smaller rings. The edge state resonances then simply the longitudinal modes of this super-ring resonator. So when the pump beam is close to an edge-state resonance, in addition to the linear dispersion, the FWM process is resonantly enhanced by the edge-state super-ring resonator.

By tuning the pump frequency and pump power, Mittal et al. observed two very distinctive regimes, namely that of phase-locked Turing rolls and Nested Solitons. In the regime of Turing rolls, all resonators that lie on the edge of the lattice show the presence of multiple equidistant peaks in the rings and only a single edge mode resonance is oscillating. Remarkably, the phase of the Turing rolls in all the edge rings was locked. At higher pump powers, a regime of Nested solitons was observed. In this regime, there was a single pulse in the ring on the edge of the lattice, and also a single super-pulse in the super-ring resonator formed by the edge states. Once again, the pulse positions in the rings were phase-locked. This nested-soliton pulse would circulate around the edge of the lattice, and around defects, without losing its phase-locking. The comb spectrum in this regime showed oscillation of multiple edge modes resonances in each FSR (each copy of QAHE), and the underlying dispersion was canceled by that introduced by the Kerr nonlinearity. It is worth noting that merely exciting the edge state resonances of the lattice did not lead to the formation of nested solitons, it also required tuning the pump frequency around the edge state resonance and the pump power. As with single-ring resonators, the phase diagram (pump frequency vs pump power) of the topological frequency comb was largely dominated by a chaotic regime. Only in very narrow regimes of pump frequency and power, were these phase-locked patterns observed. It is expected that similar physics could be explored in similar coupled-resonator systems \cite{Tusnin2021} or other platforms, for example, topological circuits where it is also possible to introduce nonlinearities. Nevertheless, an experimental realization of the topological frequency comb is yet to be realized in any platform.

\section{Quantum Topological Photonics}

Due to their built-in robustness against decoherence, photonic systems are poised to play a central role in the development of quantum technologies. In addition to being the natural choice for quantum communications, photonic systems also offer a versatile platform for quantum simulations, for example, of random walks, molecular quantum dynamics, quantum-enhanced sensing, and full-scale quantum computation using measurement-based computing \cite{Gisin2007, Kok2007, OBrien2009, Ladd2010, Guzik2012}. This is facilitated by the many photonic degrees of freedom, for example, polarization, orbital angular momentum, temporal and spectral modes, etc., onto which quantum states can be encoded, manipulated, and measured. However, the key challenge in harnessing the full potential of quantum photonic systems and further diversifying their functionalities is to achieve a scalable route for quantum engineering of the various photonic degrees of freedom via large-scale integration of photonic elements on a single chip. This large-scale photonic integration is mainly hindered by the unavoidable fabrication disorder that leads to random variations in the photonic mode structure and manifests as device-to-device variations in behavior.
Following the various demonstrations of topological robustness for classical photonic systems, it is then natural to investigate if topological protection could also be used to design robust quantum photonic devices.  A number of recent theoretical and experimental works have explored such quantum topological photonic systems in various contexts \cite{Mittal2016, Rechtsman2016, Mittal2018, Blanco2018, Barik2018, Tambasco2018, Ota2019, Wang2019, Wang2019, Mittal2021, Dai2022}. One broad category of these systems has explored the extent of topological robustness in the propagation of photons carrying quantum information, for example, encoded in temporal or spatial entanglement \cite{Mittal2016, Rechtsman2016, Tambasco2018, Wang2019, Wang2019}. Propagation of entangled photons through a disordered system can, in general, lead to the loss of quantum information. In contrast, using numerical simulations, Mittal et al. \cite{Mittal2016} and Rechtsman et al. \cite{Rechtsman2016} proposed that the topological edge states can reliably carry entangled photons. We note that the quantum information in these systems is generated outside of the topological device.

The second category of quantum topological photonic systems has explored the generation of quantum states of light \cite{Mittal2018, Blanco2018}. These systems use the second or third-order optical nonlinearities of the medium and implement spontaneous parametric processes that naturally lead to the creation of photon pairs correlated in energy-time, and space-momentum.  The presence of topological edge states is then exploited as a novel and robust route to engineer the spectral or spatial correlations in generated photon pairs.  The third category seeks to interface solid-state quantum emitters, for example, quantum dots with topological photonic systems [\cite{Barik2018, Ota2019}. The inherent directionality and the robustness of topological edge states in these systems lead to chiral light-matter interactions. In the following, we review some of the experimental demonstrations of topological robustness in quantum photonic systems.

\subsection{Topological sources of quantum light}

Sources of quantum light, in particular, correlated and entangled photon pairs, have relied on spontaneous processes such as spontaneous parametric down-conversion (SPDC) and spontaneous four-wave mixing (SFWM), in optical media with $\chi^{(2)}$ or $\chi^{(3)}$ nonlinearity, respectively \cite{Eisaman2011,Boyd2003}. In these processes, one (SPDC) or two (SFWM) photons from a strong, classical pump beam annihilate and create two daughter photons, called signal and idler photons. The parametric nature of these processes indicates that no energy or momentum is transferred between the photons and the nonlinear medium, and therefore, the pump and the generated photons conserve both energy and momentum. For example, in SFWM, $2\omega_{p} = \omega_{s} + \omega_{i}$, and $2 \vec{k}_{p} = \vec{k}_{s} + \vec{k}_{i}$, where $\omega$ and $\vec{k}$ are the frequencies and the momenta of the pump (p), signal (s) or idler (i) photons.  The underlying dispersion relation $\omega$ of the photonic mode structure couples these two relations together, and eventually, leads to non-classical energy-time and position-momentum correlations in the generated photon pairs such that they are described by a two-photon wavefunction.

Implementing SFWM and SPDC on a photonic chip offers a scalable and versatile platform to generate photon pairs with engineered spectral or spatial correlations \cite{Sharping2006, Clemmen2009, Chen2011, Engin2013, Caspani2017}. In particular, on-chip quantum light sources, using SPDC or SFWM, have now been realized on a variety of material platforms, such as silicon, silicon-nitride, lithium-niobate, aluminum-nitride, etc. \cite{Moss2013, Caspani2017, Jin2014}.  A common feature of these sources is the use of a ring resonator that can resonantly enhance the strength of nonlinear interactions and lead to higher generation rates \cite{Clemmen2009, Chen2011, Engin2013}.

With the aim of further enhancing the generation of photon pairs, and simultaneously, engineering their spectral and temporal correlations in a topologically robust way, Mittal et al. \cite{Mittal2018} used the system of coupled silicon ring resonators to implement SFWM. As we discussed earlier, this system realizes a synthetic magnetic field, and thereby, simulates the integer quantum Hall effect for photons \cite{Hafezi2011, Hafezi2013, Mittal2016}.  They chose the synthetic magnetic field flux $\phi = \pi/2$ such that the transmission spectrum of the device exhibits two edge bands, with edge states circulating around the lattice in clockwise and counter-clockwise directions. Using transmission and delay measurements made over a number of devices, the edge states in this system have been shown to be quantitatively robust against common fabrication disorders, for example, a mismatch in the ring resonance frequencies \cite{Mittal2014}.

\begin{figure*}
 \centering
 \includegraphics[width=0.7\textwidth]{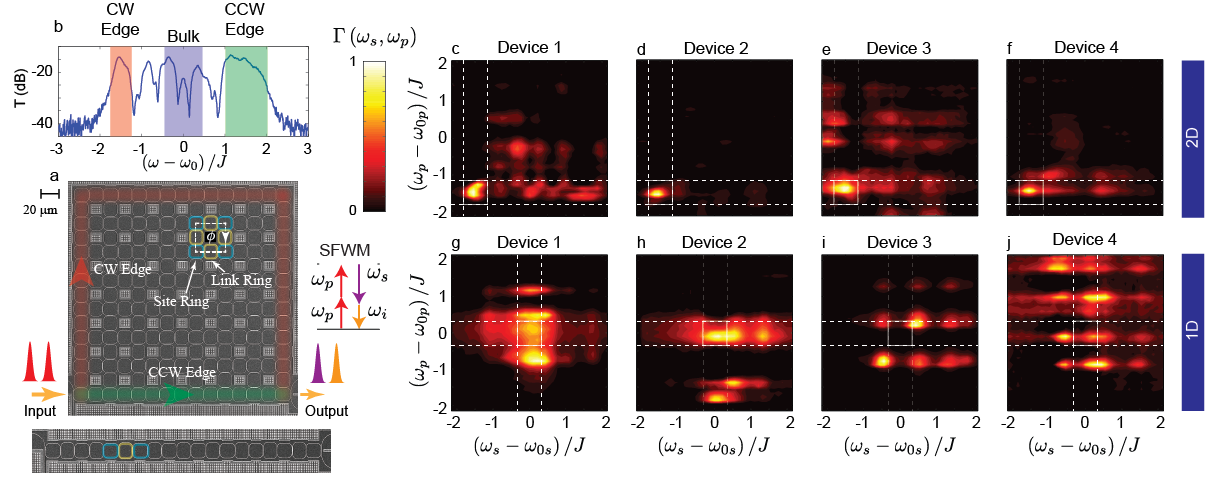}
 \caption{\textbf(a), 2D ring resonator array used to realize a topological source of correlated photon pairs generated via SFWM. Because of the synthetic magnetic field, photons acquire a non-zero, direction-dependent phase $\phi$ when they circulate around a closed path of four site rings (cyan) and four links (yellow) rings. The clockwise (CW) and the counter-clockwise (CCW) edge states are highlighted in color. \textbf(b), Measured transmission spectrum showing edge and bulk bands. \textbf(c-f)   Measured spectral correlations, that is, the number of photons generated as a function of the pump and signal frequencies. The dashed lines indicate the edge band region. The spectral correlations for 2D topological devices are very similar in the edge band region. \textbf(g) SEM image of a topologically trivial 1D array of coupled ring resonators. \textbf(h-k) Measured spectral correlations for 1D devices. The correlations differ significantly across devices because of disorder.}
 \label{fig:quantum_source}
\end{figure*}

While the topological robustness of transmission through photonic edge states has been extensively explored for applications in integrated photonic devices, Mittal et al. exploited the linear dispersion of the edge states to engineer the spectral correlations of generated photons. In particular, the spectral correlations between generated photon pairs, as well as their generation rate dictated mainly by the phase matching between the pump, the signal, and the idler photons, that is, $2 \vec{k}_{p} \left(\omega_{p}\right) = \vec{k}_{s} \left(\omega_{s}\right) + \vec{k}_{i} \left(\omega_{i}\right)$. To understand these spectral correlations, they measured the generation rate of photons as a function of the input pump frequency and the spectra of generated signal and idler photons Fig. \ref{fig:quantum_source}c-f. They showed that the maximum number of photons is generated when the pump frequency is in the edge band of the device (highlighted by the white box). Furthermore, this also limits the spectra of generated photon pairs to the same edge band. This spectrally confined and enhanced generation of photon pairs is because of the linear dispersion of the edge states, which naturally satisfies the phase-matching condition when all the photon fields are in the edge band of the device. Furthermore, their confinement at the edge of the lattice ensures that they also have an excellent spatial overlap and enhanced the generation of photon pairs.  In contrast to the edge modes, the bulk modes show a much weaker generation of photon pairs with no spectral confinement.

To test the robustness of spectral correlations between photons generated by their topological source,  Mittal et al., made measurements over a number of devices, and also compared their results against a similar source implemented using a topologically-trivial one-dimensional array of ring resonators (Fig. \ref{fig:quantum_source}a) \cite{Davanco2012,Kumar2014}. Although these devices were fabricated at state-of-the-art commercial silicon foundries, they had a significant disorder in the ring resonance frequencies, hopping strengths, as well as hopping phases, \cite{Mittal2014}. Nevertheless, as expected, they observed that for topological sources, the maximum number of photons was always generated when the pump, the signal, and the idler fields constituted the edge modes of the device, and therefore, the spectral correlations in the edge band were very similar across different devices. In contrast, the topologically-trivial 1D sources showed very significant variations in their correlations, with a much lower similarity across devices. Using second-order cross- and self-correlation measurements between generated photons, they confirmed that their source was operating in the quantum regime. More recently, this scheme has also been extended to generate path-entangled photon pairs \cite{Mittal2021, Dai2022}. These results bode well for the use of topological sources to achieve quantum interference between photons generated by independent sources.

In another similar experiment, Blanco et al. \cite{Blanco2018} investigated the generation of correlated photon pairs in a 1D lattice of coupled waveguides. The coupling strength between the waveguides is modulated by alternating the gap (small or large) between the waveguides, such that the lattice simulates the SSH model \cite{Ozawa2019}, and the edge states appear at the physical boundary between the two topological phases. Furthermore, the edge state wavefunction vanishes at lattice sites immediately neighboring the edge site and alternating waveguides thereafter.

The waveguides were fabricated using silicon which allowed for the generation of correlated photon pairs via SFWM.  In particular, the edge state of the lattice was pumped using a pulsed laser that generated signal and idler photon pairs as it propagated through the lattice. At the output of the lattice, Blanco et al. measured the spatial correlations in the generated signal and idler photons. They observed that similar to the classical (single-particle) edge state wavefunction in the SSH model, the spatial correlations in the two-photon wavefunction also showed zero amplitude at alternating waveguides. Furthermore, they showed that these zeros of the wavefunction were robust against disorder in the coupling strength between the waveguides.

From a more fundamental perspective, the SFWM process coherently adds or removes photon pairs from the topological lattice. The number of particles in the lattice is, therefore, not conserved and it operates in the non-Hermitian regime, with no analogs in fermionic systems. Furthermore, as theoretically shown by Peano et al. \cite{Peano2016}, an increase in the strength of the SFWM interaction would naturally lead to the generation of squeezed light such that only the topological edge modes of the lattice are effectively squeezed.

\subsection{Topological robustness for propagating quantum states of light}

While photons do not interact with one another, they do exhibit quantum interference which forms the basis of many algorithms used in quantum communications, quantum simulations and quantum computation using photons \cite{Kok2007, OBrien2009, Ladd2010}. This is best exemplified by the Hong-Ou-Mandel interference where two indistinguishable photons arriving in two different input ports of a beam-splitter tend to bunch at either of the output ports \cite{Hong1987}. This interference phenomenon has led to the observation of quantum walks of correlated photons and the realization of boson sampling in spatial networks of integrated beam-splitters \cite{Guzik2012, Pan2012}.

However, scaling this multi-photon quantum interference and boson sampling schemes to a larger number of photons requires a significant reduction in the variations of the splitting ratio of the on-chip beam-splitters. It is therefore natural to investigate if topological protection can be used to design robust beam splitters.

Along these lines, Tambasco et al. \cite{Tambasco2018} realized a beam-splitter using the topological edge modes. Their system consists of 1D arrays of coupled waveguides, such that the coupling strength between them is modulated both along the lattice and along the length of the waveguides. This system simulates the off-diagonal Harper model and hosts a pair of edge states at its boundaries, similar to the SSH model. However, modulation of the coupling strength along the length of the waveguides allows them to adiabatically delocalize the edge states from the boundary to the bulk of the lattice such that the photons traveling in the edge states can now interfere. The edge modes are then again localized at the boundaries of the lattice. This setup then realizes an integrated beam-splitter for photons but uses edge modes for guiding photons.

At the input of this topological beam-splitter, Tambasco et al. injected two indistinguishable photons generated via off-chip SPDC. By tuning the relative delay between the input photons and using coincidence measurements at the output, they observed a high visibility HOM interference dip, which confirmed the intended operation of their topological beam-splitter.  However, the robustness of the beam-splitting ratio of this topological beam-splitter against fabrication disorder is yet to be studied.

In another experiment using similar 1D arrays of waveguides that simulates the off-diagonal Harper model, Wang et al. \cite{Wang2019} investigated the robustness of intensity correlations between indistinguishable photon pairs as they propagate through the lattice. Similar to the experiment of Tambasco et al., the correlated photon pairs were generated off-chip using SPDC. Wang et al. showed that when both the photons are injected in the edge mode of the array, they maintain the intensity correlations. In contrast, when the photon pairs propagate through bulk modes, there is a suppression in their correlations.

In a similar context, Mittal et al. \cite{Mittal2016} numerically studied the propagation of time-bin entangled photons through their 2D topological system of coupled ring resonators. Similarly, Rechtsman et al. \cite{Rechtsman2016} investigated the propagation of spatially entangled photon pairs through their Floquet topological system of coupled helical waveguides. These investigations are similar in essence to quantum walks of photon pairs through networks of beam-splitters or coupled waveguides. They observed that propagation through edge states preserves the temporal and spatial correlation between photon pairs, respectively, even in the presence of disorder.

\subsection{Topological photonic systems coupled to quantum emitters}
Coupling light to matter degrees of freedom, such as, quantum dots, can mediate the interaction between photons and lead to novel quantum states of light \cite{Lodahl2015}. In turn, the photonic mode structure can significantly alter the properties of solid-state systems. For example, photonic cavities can be used to manipulate the emission spectra and the excitation lifetimes in quantum dots \cite{Shields2007,Lodahl2015}. Coupling quantum dots and solid-state emitters to topological photonic edge states is, therefore, an exciting avenue to investigate chiral light-matter interactions that could lead to many-body states \cite{Pichler2017}.

\begin{figure*}
 \centering
 \includegraphics[width=0.7\textwidth]{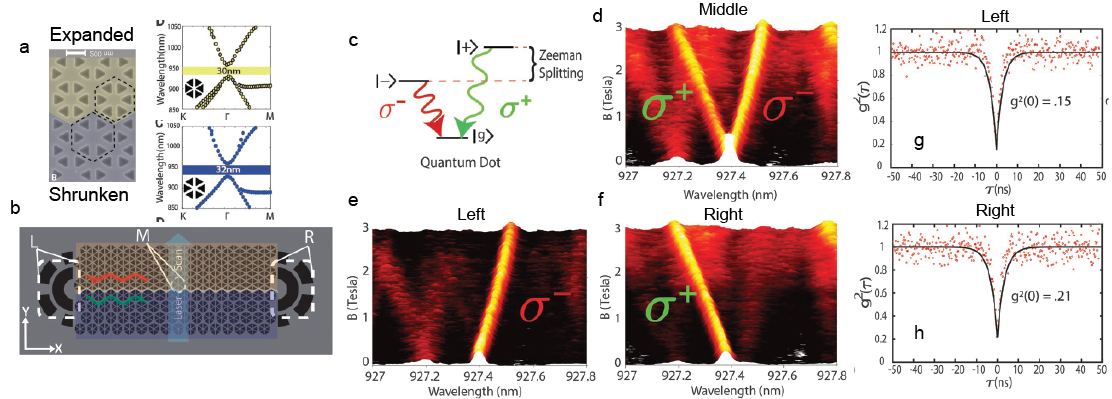}
 \caption{\textbf(a-c) SEM image, and the band structure of the shrunken and expanded honeycomb lattice used to realize a topological interface between quantum dots and helical edge states. (c) An applied magnetic field introduces a Zeeman splitting between the pseudo-spin (right and left-circular) polarized photons.  (d-f) The pseudo-spin polarized edge states propagate along the interface in opposite directions, where they are collected using grating couplers. (g,h) The measured second-order correlation function $g_{2}\left(\tau \right)$ shows the operation of quantum dots as single photon sources. \cite{Barik2020}.}
 \label{fig:TPCQD}
\end{figure*}

Barik et al. \cite{Barik2018} realized such a quantum optics interface between quantum dots and photonic edge states. Their topological photonic system was designed using a 2D photonic crystal with triangular holes in a GaAs membrane Fig. \ref{fig:TPCQD}b. When the holes are arranged in a honeycomb lattice, the band structure of the photonic crystal exhibits a Dirac point, very similar to that of graphene \cite{Wu2015, Barik2016}. Nevertheless, a deformation of the unit cell of the lattice leads to the appearance of a bandgap. More specifically, expanding the unit cell of the lattice, that is, increasing the distance between the holes in the unit cell while keeping the boundaries of the unit cell constant, resulted in a bandgap that was topological in nature. In contrast, shrinking the unit cell also opened a bandgap, but a trivial one. Therefore, an interface between the shrunken and the expanded domains hosts topological edge states. This model realizes the quantum spin Hall effect where the in-plane circular polarization of the electric field constitutes two pseudo-spins of the system. The edge states corresponding to the two pseudo spins propagate along the interface in opposite directions.

The GaAs membrane used by \cite{Barik2018} was embedded with InAs quantum dots, with their emission spectra well aligned to the bandgap of the photonic crystal structure \cite{Barik2018}. To couple the quantum dots to the in-plane circularly polarized photonic edge states, they used an out-of-plane magnetic field that induced a Zeeman splitting in the excited state energies of the quantum dots (Fig. \ref{fig:TPCQD}d). In this configuration, the two Zeeman-split energy levels were selectively coupled to the two circularly polarized photonic edge states propagating along opposite directions. In this work, \cite{Barik2018}, Barik et al. excited a single quantum dot in the middle (M) of the interface (see Fig. \ref{fig:TPCQD}b) and measured the spectra of photons collected from either side (L or R) of the interface, as a function of the magnetic field strength. Because of the pseudo-spin selective coupling of the excited states to counter-propagating edge states, they observed that the lower wavelength (higher energy) photons were primarily guided by the edge states to the right end of the interface whereas the higher wavelength (lower energy) photons were guided towards the left end of the interface. To demonstrate the robustness of this topological quantum-optics interface, they showed that the chiral propagation of photons is robust against any disorders that do not flip the two pseudo-spins, for example, bends in the interface. Furthermore, they used second-order correlation measurements to observe the anti-bunching of photons which ensured that they were indeed single photons.

In another experiment, Ota et al. \cite{Ota2019} explored the coupling of quantum dots to nanophotonic cavities realized using corner states of light in higher-order topological systems. Similar to Barik et al., their system comprised a GaAs photonic crystal membrane, with embedded InAs quantum dots. The photonic crystal was designed by etching two sets of square holes, with different lengths of their sides, in the GaAs membrane.  This difference in the hole dimensions opened up a bandgap. More importantly, a $90^\circ$ interface between two photonic crystal regions with swapped hole dimensions led to the emergence of corner states in the bandgap, physically located at the bend in the interface. This system is analogous to a two-dimensional SSH model with alternating coupling strengths in both dimensions. To probe the existence of corner states in their photonic crystal, Ota et al. used the photoluminescence from the ensemble of quantum dots as a broadband light source (pumped by a laser). As an indication of the corner states, they observed a sharp peak in the photoluminescence spectrum, in the region expected to host corner states.  They confirmed their observation of the corner states by showing that this peak in the photoluminescence spectrum originated from a narrow spatial region at the $90^\circ$ bend in the interface between topological and trivial domains of the photonic crystal. We note that unlike Barik et al. who coupled a single quantum dot to the topological edge states, Ota et al. used an ensemble of quantum dots.

\section{Remaining challenges and future directions}

Here, we highlight some of the challenges and potential future directions in topological photonics in linear, nonlinear, and quantum regimes,  as well as coupled electron-photon systems, 
 ranging  from  fundamental to application perspectives. For the latter, it is essential to go beyond proof-of-principle experiments, and a side-by-side comparison of the efficiency and yield of a topological design compared to trivial counterparts (with the same fabrication process and material) should be established. Such side-by-side comparisons and yield estimates remain scarce in the literature \cite{Mittal2014,Rosiek2023}.

\subsection{Linear topological photonics} 

\begin{itemize}
    \item In spin-Hall PhC waveguides \cite{Barik2018} and ring resonators \cite{JalaliMehrabad_APL} propagation length and Q factors are low since the edge states are above the light-cone, and therefore, radiative. Moreover, practical edge state bandwidths in spin-Hall PhC waveguides in strong perturbation (shrinking and expanding) regimes \cite{Barik2018} are limited. One intriguing improvement can be the realization of broadband spin-Hall TPC waveguides and high Q photonic cavities with below-light cone edge dispersion. Another remaining challenge is the efficiency of mode conversion at the topological-conventional waveguide interfaces \cite{JalaliMehrabad_Optica,Shalaev2018,Mehrabad2023}. Further optimization of such mode conversion is essential for the efficient integration of these optical components for scalable photonic circuitry. One recent approach to address this challenge can be found here \cite{Kagami2020}. More promising approaches may be using the recently demonstrated topological funneling of light and edge mode tapering, although it should be noted that topological funneling of light in these studies is based on non-Hermitian physics \cite{Weidemann2020,Flower2022}. Moreover, so far many valley-Hall PhC waveguides \cite{Shalaev2018,Ma2016,JalaliMehrabad_Optica} have been proposed and studied. In addition to a recent investigation of the nature and degree of backscattering against sharp bends and fabrication imperfections \cite{Rosiek2023}, further studies are required to confirm if there is any protection against real-world defects and classify and quantify the strength of such protection rigorously. A recent study can be found here \cite{Amo2023}.

    \item While helical topological waveguides have been experimentally  realized \cite{Rechtsman2016}, their photonic applications have not been explored yet. This is in contrast to coupled rings used for solitons and lasers, and TPCs used for routing and QDs. It would be desirable to explore whether similar or other exclusive applications such as chip-integrated photonic circuits are realizable using the topological helical waveguides. 

    \item One exciting direction can be the design and realization of reconfigurable topological devices with phase-changing materials. Such a reconfigurable platform can be inspired by recently demonstrated reconfigurable non-Hermitian topological photonic routing \cite{Zhao2019}. In particular, it is intriguing to be able to imprint a wide variety of optical components such as waveguides, ring resonators, and beam splitters all within the same device with a compact footprint. A recent proposal has explored the possibility of such systems in spin-Hall TPCs \cite{Tang2023}.

    \item Another potential direction in either topological ring resonator arrays or TPCs can be the realization of topological bandpass and notch filters, enabled by the robustness propagation of the edge states in larger device sizes. In particular, the realization of robust topological delay lines has been proposed in such devices \cite{Hafezi2011}. One can investigate what other devices are possible to realize that can benefit from topological protection. For example, the realization of topological photonic taper \cite{Flower2022} and application of topological photonic beaming \cite{Lee2022} has not been reported yet. Moreover, the scalability and application of topological antennas have not been investigated fully yet \cite{Lumer2020,Gorlach2018}.

    \item Spin-flip in topological photonic resonators and photonic crystal waveguides is an undesirable feature. Using inverse design or more generally machine learning techniques to address these issues and design practical topological photonic systems can be an intriguing avenue to improve device functionality and scalability \cite{chen2022inverse}. In this approach, topology can be hard coded in the optimization process. For example, a non-zero Berry phase can be encoded as a loss function.

    \item Despite decades of theoretical and experimental work on the electronic integer quantum Hall effect, the plateau transition remains an active area of debate \cite{IQHE_transitionPRB2019,ippoliti2020}. It is intriguing to explore whether photonic systems can shed light on the nature of extended states and transition between plateaus \cite{chattopadhyay2021mode}. This might require strong interaction between photons.

    \item In TPCs, valuable recent work \cite{Rosiek2023} has exhaustively investigated the nature and the degree of scattering in topological photonic waveguides. More experimental comparison is direly needed to further deepen the fundamental and technical limits of topological protection for device engineering.

    \item With recent remarkable advances in machine learning, it is interesting to investigate its implication in physics \cite{Giuseppe2019}, and more specifically in topological photonic systems. This can be in diagnosing and classifying topological states and even finding new band structures with topological phases \cite{ma2023topogivity}. In addition to designing topological photonics devices, exploring the connection between topological data analysis (TDA) and machine learning and their potential implications in photonic systems may also be another emerging avenue in this field \cite{leykam2023topological}.
    
    \item Another novel direction can be the realization of reconfigurable topological photonic devices using phase-changing materials. In particular, the demonstration of multi-functional operation within the same photonic device footprint can be intriguing. A recent example of such a study can be found here \cite{tian2023tunable}.
    
\end{itemize}

\subsection{Nonlinear topological photonics}

\begin{itemize}
    \item Recent theoretical studies suggest a rich set of phenomena to occur in topological lasers \cite{Amelio2020,Loirette-Pelous2021}. Experimental exploration of such ideas can significantly expand the use of topological edge states in real-world applications in novel lasers. Experimentally, an unambiguous demonstration of topological robustness in topological lasers and a comparison of their efficiency compared to trivial one-dimensional counterparts in a side-by-side comparison remains an active area of research. Another direction includes Dirac-point lasers in 2D geometries \cite{Chua14,Gao2020,Yang2022}. In particular,  it is useful to optimize the stability and laser emission in broader chip areas without multi-mode operations in these devices. Moreover, Weyl-points in 3D photonic crystals are novel candidates for expanding topological lasers to more than 2D configurations \cite{Sachin2020,Christina2022}. Polaritons, which are hybrid photon-exciton particles, and either photon, excitons, or their coupling form can also be engineered to have topological properties. In these topological exciton-Polaritons, which were demonstrated recently (\cite{Klembt2018}), the band gap is very small, which makes their broadband application and spectrally-resolved demonstration challenging. It would also be intriguing to explore concepts such as spin-selective strong light-matter interaction in topological exciton-polariton systems \cite{suarez2023spin}.

    \item Recently Topological frequency combs and nested temporal solitons have been theoretically proposed \cite{Mittal2021}. Nevertheless, the experimental realization of topological optical frequency combs using coupled ring resonators is expected to be challenging. In particular, the currently estimated pump power requirement for topological frequency combs is high, more than 10W. This is mainly set by the disorder in ring resonance frequencies which, even for state-of-the-art photonic integration, is of the order of a few tens of GHz. This sets a lower limit on the coupling strength $J$ between the rings and limits the loaded quality factor of the rings. Furthermore, lowering pump power requirements will also reduce the deleterious thermal effects which are problematic even for single-ring combs. Another area of concern for ring-resonator-based topological combs will be the mode-mixing between transverse modes of the ring waveguides. To lower losses, single-ring resonator frequency combs often employ waveguides that support multiple transverse modes. However, for coupled ring resonators, mixing between different transverse modes can be a significant challenge. For proof-of-principle demonstration, many of these issues could be mitigated by designing rings with lower coupling strengths (higher loaded quality factors and lower topological edge bandwidth) and coarse-tuning the ring resonator frequencies (say using heaters) such that the disorder falls within the reduced topological edge bandwidth. Nevertheless, it will be interesting to investigate other topological photonic designs that could lead to lower pump power requirements and make them more appealing for practical applications. From a theoretical perspective, the topological frequency combs could host a much more diverse range of nonlinear solutions, such as breathing solitons, dark solitons, platicons, etc.; these solutions have not yet been explored. The development of an analytical approach to describe these multi-resonator systems might be very helpful, but again, it is expected to be challenging.

    \item While the nature of the linear and quantum many-body topological states has been heavily studied and understood, the nature of topological invariants in the nonlinear topological photonic systems remains elusive. Recent works have shown that in the nonlinear regime, the topological character of the linear regime is inherited in some form \cite{Ezawa2022,Mukherjee2021,Mostaan2022}. It is interesting to see whether there are genuinely nonlinear topological states. Moreover, is it possible to have a bulk-edge correspondence for the nonlinear topological states? 

    \item Inspired by the above examples, are there other photonic phenomena without electronic counterparts? An example can be using topological confinement for more efficient lasers \cite{Shao2019}. In particular, it is useful to answer whether there are other implications for using the same confinement, such as optical sensing. Moreover, further investigation of the topological amplifier which was recently reported \cite{Sohn2022} is also another potential future direction. 
\end{itemize}

\subsection{Quantum topological photonics}

\begin{itemize}
    \item In pair generation in a topological lattice, the key advantage is the robust phase-matching compared to conventional counterparts \cite{Mittal2018}. It would be intriguing to find more applications of this robust phase-matching advantage for the realization of non-linear quantum optical effects. Moreover, the device footprint in a topological quantum light generation device is considerably large since it is comprised of rings with several hundreds of microns.

    \item Position-dependence of chiral coupling in QD-coupled topological waveguides is a significant limitation in these quantum optics interfaces. Moreover, chirality and high Beta-factor areas are mostly present in the holes rather than the material \cite{Nussbaum2022,JalaliMehrabad_Optica,martin2023topological}, which is detrimental for coupling to solid-state quantum emitters. Also, low Purcell factor (currently only up to less than 5 is reported \cite{Barik2020}) is another limitation in QD-coupled TPCs. In topological waveguides, the emitter's coupling efficiency, as well as emission enhancement, may be improved with either slow-light effect \cite{Arakawa2022} or smaller mode volume (for example harnessing topological mode tapering \cite{Flower2022}). Moreover, in whispering-galley mode ring resonators, \cite{Barik2020,JalaliMehrabad_Optica}, possibilities for achieving higher Q factors (for example using surface passivation for suppression of the out-of-plane scattering) can be investigated. Similar to \cite{Grim2019}, coupling multiple quantum dots to edge states can be a very interesting direction to explore, in particular of interest is the collective dynamics between distance emitters \cite{lodhal2023}, however, this is currently challenging due to the beta factor being position-dependent in current TPC waveguides. Chiral coupling is also position dependent in these systems so the directional sub and superradiance effects from embedded quantum emitters in topological waveguides are challenging. One avenue to explore can be using inverse design to address this issue, and in turn, investigate if such a platform can be utilized to address the spatial inhomogeneity challenge in chip-integrated solid-state quantum emitters. Finally, the realization of on-chip quantum interference of single photons in add-drop filter photonic crystal configurations, in which chiral coupling between emitters and several modes of the resonator was shown recently \cite{Mehrabad2023}, would be another intriguing possibility to explore. In particular, combining recent demonstrations of broadband slow-light enhancement in a topological photonic ring resonator with an integrated add-drop filter configuration may be studied \cite{Xie2021,Mehrabad2023}.

    \item Another interesting development is the possibility of coupling emitters to topological photonic structures. This includes coupling emitters to 1D structures\cite{bello2019unconventional,Leonforte2021, kim2021quantum}, and 2D systems\cite{de2021light,vega2023topological} with novel forms light-matter hybrids.
    
    \item An extremely exciting avenue is the realization of Laughlin states with a higher photon number than two \cite{Schine2016}, and more broadly, other topologically-ordered states and potentially braiding  them \cite{satzinger2021realizing,google2023non}. Specifically, the excitations above the ground states of such models can have \emph{anyonic statistics} this is neither fermionic nor bosonic \cite{Wen2017}. The non-Abelian anyons have been proposed as a robust scheme for topological quantum computation \cite{Sankar2008}. Note that one can simulate such exotic statistics even with non-interacting photons \cite{noh2020braiding,yang2019synthesis,zhang2022non}. However, the many-body features such as the topological robustness for quantum computation are absent in such non-interacting systems.
\end{itemize}

\subsection{Strong photon-photon interaction and coupled electron-photon systems}

If the interaction between cavity photons is so strong that a single photon can prevent the transmission of another one (photon blockade), one can expect even more exotic topological states, such as the photonic counterpart  of fractional quantum Hall states. In particular, there is a whole class of topological states, known as topologically ordered states, which are distinct from the states covered in the introduction. In these states, entanglement and strong interaction play a central role, for which a brief review can be found here \cite{Wen2017}. In fact, once photons strongly interact with each other, they can be considered as spin-1/2 particles and therefore many of the topologically ordered models will be directly applicable. For the case of photonic fractional quantum Hall states, the essential ingredients are gauge fields (discussed earlier) and strong photon-photon interaction. The important parameter in such systems is the magnetic filling factor: the number of particles divided by the number of magnetic flux (introduced in Sec.\ref{sec:semiclassical}). The simplest case for bosons is $\nu=1/2$, where the ground state is Laughlin state, and it is both unique and gapped.  For a pedagogical review of electric and boson fractional quantum Hall refer to \cite{Girvin1999,cooper2008rapidly}, respectively. 

Generically in such systems, the ground state on a torus (periodic boundary condition) has a finite degeneracy. Therefore, a Chern number can not be associated with a single state, and indeed it is \textit{shared} among the degenerate states, and therefore, it can be fractional. 

The strong interaction could be achieved in various ways such as Rydberg atoms or superconducting qubits \cite{carusotto2020photonic}. Remarkably, in the Rydberg systems, fractional quantum Hall states (Laughlin states) of a few photons have been observed \cite{Clark2020}. Scaling such a system to a larger number of particles remains a challenge. In fact, one may ask how small of a system one can call a topologically ordered matter. In other words, given a wavefunction on finite system size, is it possible to identify whether a system is topologically ordered or not? Can one extract the Chern number, without prior knowledge of the Hamiltonian and application of any field? The answer to these questions seems to be positive based on recent analytical and numerical works \cite{dehghani2021extraction,cian2021many,fan2022extracting}, however, there is no experimental demonstration to this date.

So far we considered purely photonic models, whereby by engineering a Hamiltonian with topological properties, directly for photons, one can observe various topological phenomena. An interesting direction is to consider light-matter coupling where either the photonic or matter part has some topological properties, and therefore, the coupled system inherits those topological features. In other words, the matter part is not \textit{integrated out} and the light-matter interplay is the essence.

We note that the above categorization might sometimes seem artificial since the underlying microscopic theory for all the cases in this Perspective is quantum electrodynamic. Specifically, what is purely photonic or matter is a matter of length scale over which we integrate out microscopic degrees of freedom to write an effective Hamiltonian for the system. For example, one can call quantum Hall states coupled to optical cavities also a topopolariton since the optical excitations in the cavity-quantum Hall system can be considered excitons that are coupled to the cavities. Below, we highlight several directions. 

\begin{itemize}

\item{Light-matter interaction in electronic quantum Hall systems:}
As we mentioned at the beginning of this Perspective, electronic quantum Hall systems are the first physical systems to manifest topological properties in transport measurements. However, from early on optical measurements were also performed on such systems \cite{sarma2008perspectives},  for example, to probe electronic incompressibility. More recently, there have been interacting experiments to couple such states to cavities, either in the THz, \cite{appugliese2022breakdown,ciuti2021cavity} or optical domain \cite{Knuppel2019nonlinear}, to probe and manipulate intra- and interband states, respectively. 

Regardless of being in the cavity or free space, it is intriguing to ask whether light-matter coupling could be exploited to create and manipulate electronic topological states, and eventually perform braiding. 

Moreover, it has been theoretically argued that the light-matter interaction is dramatically modified in quantum Hall states, since the chirality and topological robustness of the electronic states may lead to the spatially large wave functions, which could be comparable to the corresponding optical transitions \cite{Gullans2017multipole}. In particular, the dipole approximation can be violated and the system could be sensitive to the gradient of the electric field and the phase of an optical vortex beam. The latter is only possible if the electron is phase coherent around the optical vortex and \textit{experiences} the phase winding of the optical beam. It has been proposed that such light-matter coupling could lead to radial current in quantum Hall systems in the absence of any electric field bias \cite{cao2021optical, Fleischhaue2018}. Such optical vortex beams could be used to optically create topological excitation in fractional quantum Hall systems \cite{grass2020optical}. A recent experiment demonstrated that photocurrent could be sensitive to the beam phase winding \cite{Ji2020}. In the context of this section, these systems are particularly interesting because both the electronic and photonic states have topological properties and such topological interplay is an interesting direction of future research.

\item{Topological photonic crystals:}  In the linear section, we discussed photonic crystals that can have topological properties, manifested in the presence of helical waveguides and their coupling to point-like emitters, like QDs \cite{martin2023topological}. One can also couple extended excitons states, such as the ones in 2D materials with optical transition, to such helical states. Since layered 2D materials are essentially one or a few atomic layers thick, they can strongly couple to the confined electromagnetic modes of the topological photonic crystals. Particularly interesting are recent studies on the hybridization of topological photonics states to condensed matter systems, where 2D transition metal dichalcogenides were shown to be strongly coupled to topological photonic crystal metasurfaces, forming a polaritonic metasurface \cite{Li2021}.

Similar to the case of quantum QDs \cite{martin2023topological}, the chiral light-matter coupling is sensitive to the location of the emitter with respect to the transverse position of the waveguide. In fact, one would naively expect that chiral light-matter coupling for 2D material excitons would be absent in such systems, because excitons, with the same polarization, are present all along the transverse direction of the waveguide, half coupled to the left-propagating modes and half to the right-propagating modes. However, experimental observation does not agree with this argument, and this subject remains an active area of research.

\end{itemize}

\begin{acknowledgments}
We wish to gratefully acknowledge Erin Knutson, Daniel Leykam, Mikael Rechtsman, and Alberto Amo for their insightful comments and discussions during the preparation of this manuscript. This work was supported by AFOSR FA9550-20-1-0223, FA9550-19-1-0399, and ONR N00014-20-1-2325, NSF IMOD DMR-2019444, ARL W911NF1920181, Minta Martin and Simons Foundations.
\end{acknowledgments}

\bibliography{Main}
\end{document}